\pdfoutput=1
\documentclass[12pt]{article}
\usepackage[margin=1in]{geometry}
\usepackage{amsmath}
\usepackage{titling}
\usepackage{blindtext}
\usepackage{pgfplots}
\usepackage{natbib}
\usepackage{comment}
\usepackage{setspace}
\usepackage{authblk}
\usepackage[percent]{overpic}
\usepackage[section]{placeins}
\usepackage{pbox}
\usepackage{comment}
\usepackage{titling}
\usepackage{natbib}
\usepackage{soul}
\usepackage{siunitx}
\usepackage{textcomp}
\newcommand{\GG}[1]{}

\usepackage{ragged2e}
\justifying

\newcommand{\gtsim}{\raisebox{-1.0ex}{$\stackrel{\textstyle>}{\sim}$}}
\newcommand{\ltsim}{\raisebox{-1.0ex}{$\stackrel{\textstyle<}{\sim}$}}

\def\yohkoh{{\sl Yohkoh}}

\def\hinode{{\sl Hinode}}

\def\p78{{\sl P78-1}}

\def\sdo{{\sl SDO}}

\def\halpha{H$\alpha$}

\begin{document}

\title{The Missing Cool Corona in the Flat Magnetic Field around Solar Active Regions}

\author[1]{Talwinder Singh}
\affil[1]{Department of Space Science, The University of Alabama in Huntsville, AL 35805, USA}

\author[2]{Alphonse C. Sterling}
\affil[2]{NASA/Marshall Space Flight Center, Huntsville, AL, 35806, USA}

\author[2,3]{Ronald L. Moore}
\affil[3]{Center for Space Plasma and Aeronomic Research, The University of Alabama in Huntsville, AL 35805, USA}

\setcounter{Maxaffil}{0}
\renewcommand\Affilfont{\itshape\small}
\date{}  

\begin{titlingpage}
    \maketitle

\begin{abstract}

SDO/AIA images the full solar disk in several EUV bands that are each sensitive to coronal
plasma emissions of one or more specific temperatures. We observe that when isolated active
regions (ARs) are on the disk, full-disk images in some of the coronal EUV channels show the
outskirts of the AR as a dark moat surrounding the AR.  Here we present seven specific
examples, selected from time periods when there was only a single AR present on the
disk.  Visually, we observe the moat to be most prominent in the AIA 171~\AA\ band, which
has the most sensitivity to emission from plasma at  $\textup{log}_{10} T = 5.8$.  By examining the 1D
line-of-sight emission measure temperature distribution found from six AIA EUV channels, we find the
intensity of the moat to be most depressed over the temperature range
$\textup{log}_{10} T \approx 5.7 - 6.2$ for most of the cases.  We argue that the
dark moat exists because the pressure from the strong magnetic field that splays out from
the AR presses down on underlying magnetic loops, flattening those loops -- along with
the lowest of the AR's own loops over the moat -- to a low altitude.  Those loops, which would normally
emit the bulk of the 171~{\AA} emission, are restricted to heights above the surface
that are too low to have 171~{\AA}-emitting plasmas sustained in them, according to
\citet{Antiochos86}, while hotter EUV-emitting plasmas are sustained in the overlying
higher-altitude long AR-rooted coronal loops. This potentially explains the
low-coronal-temperature dark moats surrounding the ARs.

\end{abstract}
\end{titlingpage}

\section{Introduction}

The Sun's outer atmosphere, the corona, resides beyond the 6000~K solar surface and extends into
the inner heliosphere.  It is characterized by temperatures near and upward of $10^6$~K, and hence
radiates strongly in the extreme ultraviolet (EUV)  and X-ray portion of the  spectrum. 
Photographs in broad-band X-rays, such as from {\it Skylab} in the 1970s \citep{rosner.et78} and later from X-ray
telescopes on the \yohkoh\ \citep{Ogawara91} and \hinode\ \citep{Kosugi07} missions, show three prominent large-scale morphological
coronal regions. First there  is the general roughly uniform glow characteristic of quiet
regions, second are localized comparatively  bright locations in active regions (ARs), and
third are the regions that appear dark in the X-ray images and hence are known as coronal holes.  Often the
coronal holes are localized around the polar regions (polar coronal holes), but they also can
extend down to lower latitudes from the polar regions, or even develop at lower latitudes isolated
from the poles (equatorial coronal holes) \textup{\citep{Harvey02}}.

This overall coronal morphology reflects the strength and  topology of the magnetic field in the various
regions: In ARs, the field is stronger and consists largely of dynamic loops, upon which coronal
heating is obviously strong,  resulting in relatively high temperatures
(few times $10^6$~K) and high intensity in X-rays and EUV\@ \textup{\citep{Warren03,Winebarger03}}. The corona in quiet regions
consists largely of weaker-field closed magnetic loops, including some of a size scale larger than that
of ARs.  Their temperature ($\sim$10$^6$~K) is lower than that of coronal loops in ARs due to their weaker heating resulting from the weaker field
strength and less-dynamic magnetic activity at their photospheric roots.   Finally,
coronal holes in contrast occur where the magnetic field is of quiet-region strength but is largely unipolar, and is ``open" in the  sense that
it extends far into the heliosphere, with a relatively low coronal temperature because
the coronal plasma and its thermal energy are efficiently lost to the solar wind.  \citep[See, e.g.][for discussions of
coronal temperature and structure.]{withbroe.et77}

The Atmospheric Imaging Assembly (AIA) \citep{lemen.et12} onboard the Solar Dynamics Observatory (\sdo) \citep{Pesnell12} observes the
corona in seven different EUV channels. The passband for each of the channels however is narrower
than the above-discussed broadband X-ray images, with  the result being that  each channel tends to
sample a different temperature regime of the corona, with the channels and their respective
approximate peak coronal $\log_{10}$ temperature values being \citep{lemen.et12} 304~\AA\ (4.7),
171~\AA\ (5.8), 193~\AA\ (6.2), 211~\AA\ (6.3), 131~\AA\ (5.6), 94~\AA\ (6.8), and 335~\AA\ (6.4).
(There are however additional peaks, and overlaps in the response curves; see
\citet{lemen.et12} and \citet{Boerner2012} for details.)

\begin{figure}[!htb]
%\vspace{11pc}
\center
 
\includegraphics[scale=0.1,angle=0,width=10cm,keepaspectratio]{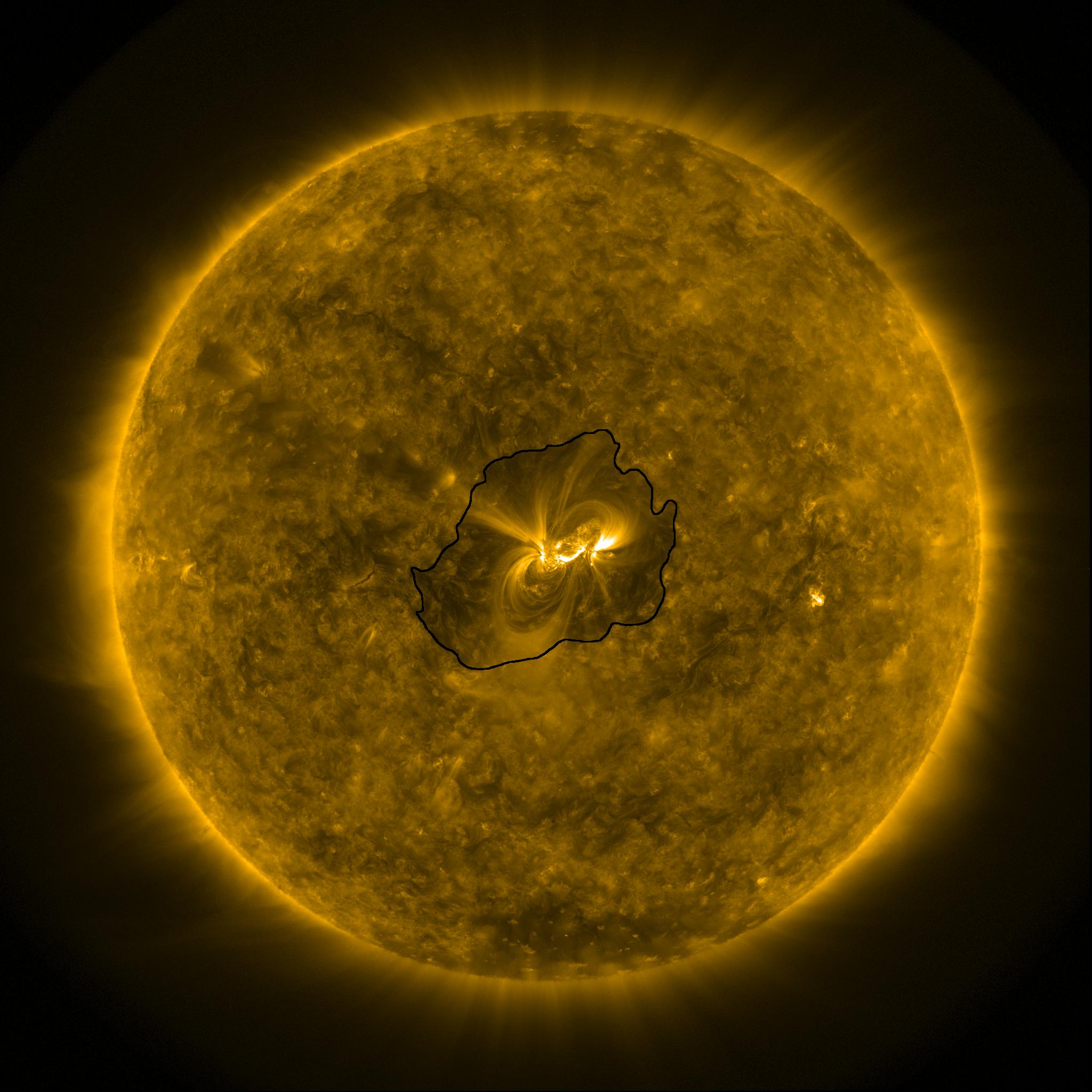}

\caption{An \sdo/AIA 171~{\AA} image from 2018 Feb 11 at 06:04~UT; case~1 of Table~1. 
The bright region near disk center is active region NOAA AR~12699.  A roughly circular moat-like annulus of 
low emission clearly surrounds the active region.  North is up and west is to the right in this and
in all other solar images in this paper. \textup{We have visually drawn a black boundary surrounding this dark region.}}
\label{Without_PFSS}
\end{figure}

AIA regularly images the full-disk Sun, which allows for unprecedented inspection of the
large-scale corona in different temperature regimes.   We observed that in some channels,
with the primary one seeming to be 171~\AA, that ARs on the visible disk frequently are
surrounded by an annular darker moat-like region, with the AR emitting brightly interior
to that annulus, and with the quiet Sun emitting with brightness typical of the quiet Sun
in regions exterior to that annulus.  Figure~\ref{Without_PFSS} shows an example from
AIA~171~\AA, where the AR is near disk center.  Although we have not carried out a
quantitative survey of their frequency,  \textup{our} inspection of \textup{AIA} images shows such dark
moats around ARs to be common, often being visible as outskirts of ARs \textup{throughout the solar cycle}.  Often the moats completely
surround the ARs, while sometimes they appear to be in more restricted locations, where
perhaps however the brighter parts of those regions are due to bright loops of the  AR
blocking the view of the moat  along the line-of-sight through the bright loops.  Here we
will attempt to  understand the nature of these moats by studying seven
examples, at times when there was a single AR on the Sun, and when that AR was located near disk center.

These moats are distinct from coronal holes, which tend to be dark across the six hotter 
AIA EUV wavelength passbands \citep[e.g.][]{garton.et18}; that includes all seven AIA EUV bands except for 304 A, which mostly shows the lower cooler transition region, \textup{the intensity 
of which is not as strongly affected by the presence of coronal holes as some of the hotter
channels \citep[e.g.,][]{Hamada17}.}  In the cooler of the six hotter AIA
EUV channels (171, 193, 211, 131~\AA), the coronal holes stand out compared to the
surrounding quiet-Sun corona, which radiates brightly at those wavelengths.  In the
two hotter channels (94  and 335~\AA), the coronal holes and much of the surrounding 
quiet Sun can both be uniformly dark, since neither the coronal-hole plasma or the
quiet-Sun plasma radiates well at those wavelengths.  In contrast, as we will see below
(cf.~Fig.~\ref{case1}), the moats around  the ARs are dark primarily in 171~\AA, along
with some other wavelengths, including 304~\AA, but less so or not at all in other
wavelengths (due to overlying bright loops along the line of sight), such as 193~\AA\@.  Moreover, some of the moat area can even be brighter than
surrounding corona in some wavelengths, such as 211~\AA\ in the case of
~Fig.~\ref{case1}.  We will  examine the visibility as a function of temperature for the
seven examples we examine in the following.

\citet{wang11} studied moat regions like ours in AIA 171~\AA\ images and line-of-sight
magnetograms; they also described them as moat-like, but primarily referred to them as
``dark canopies" around ARs.   They pointed out that the chromospheric counterpart of these regions are populated by dark \halpha\ fibrils  aligned with AR magnetic field that
have been studied by a number of authors
\citep[e.g.][]{Howard1964,Foukal1971,Reardon2009}. \citet{wang11} suggested that the moat
regions consist of low-lying EUV-absorbing fibril structures that connect opposite
magnetic polarities. Here we will study in more detail the thermal properties of the dark
regions.  We will then  consider a different idea for the lack of emission from the dark
regions, based on theoretical predictions  for the thermal stability of coronal loops at low heights. We will argue that a moat forms when loops rooted in these  regions, along
with the lowest of the AR's own loops over these regions, are pushed down to low-enough
heights by the overlying strong magnetic field that loops out from the AR\@.

%can be the coronal counterparts of
%fibrils seen in the chromosphere. This, however, does not explain the presence of coronal dark 
%cavities above the chromosphere with no fibrils, even seen in the \citet{wang11} test case. Moreover, 
%\citet{wang11} did not comment on the absence of these dark regions in some other wavelengths such as 211 {\AA}.

%\begin{comment}

%\end{comment}

%In sec. \ref{data}, we explain the data used in this work. Section \ref{observations} contains 
%the observations and discussion. We finally give our conclusions in sec. \ref{conclusions}.

\section{Instruments and Data}\label{data}

We use images from \sdo/AIA from all seven EUV channels, so that we can differentiate structures 
based on their temperature \citep{Pesnell12}. AIA has  $0''.6$ pixels, and takes a full-disk image every 12~s in
each of  the EUV channels during nominal periods; see \citet{lemen.et12} for more details. We use
aia\_prep.pro in IDL's SolarSoft library to read in and calibrate AIA level~1 data to level~1.5. \textup{Level~1 data are already processed with bad-pixel removal, de-spiking and
flat-fielding. Level~1.5 data additionally have roll-correction making solar north vertical in the images, rescaling of the images to $0''.6$  pixels, and translation that puts solar disk center at the image center.}  We also  use
 line of sight (LOS) magnetograms from \sdo's Helioseismic and Magnetic Imager (HMI) 
instrument to observe the \textup{photospheric magnetic field} \citep{Schou12, Hoeksema14}.  HMI has $0''.5$ pixels, and is capable of
high cadence (45~s) although for this study we generally used only a single representative
magnetogram for each region \citep{Schou12}.  The HMI data are  read using hmi\_prep.pro in IDL, \textup{which processes the magnetograms to have correct role angle, and translates the magnetograms to put solar disk center at the magnetogram centers.}
\citet{scherrer.et12} provide more details on HMI\@.  We use full resolution 4096 $\times$ 4096
pixel data for both AIA and HMI\@.

\vspace{0.5cm}

\begin{table}

\center

\begin{tabular}{ccc}
\hline
Case \# & Date$^a$    & NOAA AR \\ \hline
1       & 2018 Feb 11 &  12699 \\ \hline
2       & 2018 Apr 25 &  12706 \\ \hline
3       & 2018 May 30 &  12712 \\ \hline
4       & 2018 Jun 17 &  12713 \\ \hline
5       & 2018 Jul 14 &  Unnumbered$^b$ \\ \hline
6       & 2019 Feb 20 &  Unnumbered$^c$ \\ \hline
7       & 2019 Apr 15 &  12704 \\ \hline
\end{tabular}
\caption{Cases discussed in this study. $^a$The time for each case is 06:04 UT. $^b$Previous Carrington rotation AR number: 12713. $^c$Previous Carrington rotation AR number: 12733.}
\label{table1st}
\end{table}

As mentioned above, we examine ARs that are isolated on the Sun, when they are near the
center of the disk.  This allows us to show clearly the moat-like regions of missing
``cool" corona around the ARs, and also allows for more accurate  AR magnetic flux
measurements in our analysis since the LOS magnetograms lose their accuracy away from the
solar disk center. Table~1 gives the dates of  observations used in this work, along with
the NOAA AR numbers of the active regions under consideration. In cases~5 and~6, the
regions have decayed to a point where they no longer have sunspots, and hence  they no
longer have a NOAA AR designation.  For those cases, we provide the AR number for the
previous rotation in the table notes.  Case~5 is the same region as case~4, but one
rotation later. The time of observations in all the cases is 06:04 UT. 
%We will discuss other AR properties listed in Table~1 in \S\ref{monopole}.

Although we find the dark moats around ARs to be a very common phenomenon during  both
solar maximum and solar minimum periods (see \S\ref{conclusions}), our selected regions
for this study are biased towards  solar minimum since it is almost impossible to find
intervals during solar maximum  periods with just one AR at the solar disk.  \textup{By looking at solar magnetogram synoptic maps for all Carrington rotations}, we found
only seven periods where the solar disk had just one major AR since \sdo's 2010 launch, 
and these are the cases we selected for study here (Table~1).

\section{Observations and Analysis}\label{observations}

%  dip of ~0.6-0.8 MK

In this section, we will show the morphology of the dark moats surrounding ARs, and also
present quantitative  analysis confirming that there is a dip in emission at cool coronal
temperatures in these regions. 

\subsection{AIA morphology}

In the previously introduced Figure~\ref{Without_PFSS}, we show an AIA 171~\AA\ image of 
the Sun for case~1 of Table~1, as observed on 2018 Feb 11 at 06:04~UT\@. The AR,  near
disk center, is NOAA region AR~12699. A roughly circular  moat of low emission can
clearly be seen around this AR\@. At this wavelength, the emission  from this dark moat
appears comparable to that from the polar coronal holes, visible at the south pole and in
the eastern part of the north pole. \textup{In fact, we found that the average emission in the moat region for this case was 96.2 DNs, whereas the average emmision in the coronal hole at the southern pole was 100.8 DNs}.

\begin{figure}[!htb]
%\vspace{11pc}
\center
\begin{tabular}{c c c c}  

\hspace{-0.6pc}
\begin{overpic}[scale=0.1,angle=0,width=4cm,keepaspectratio]{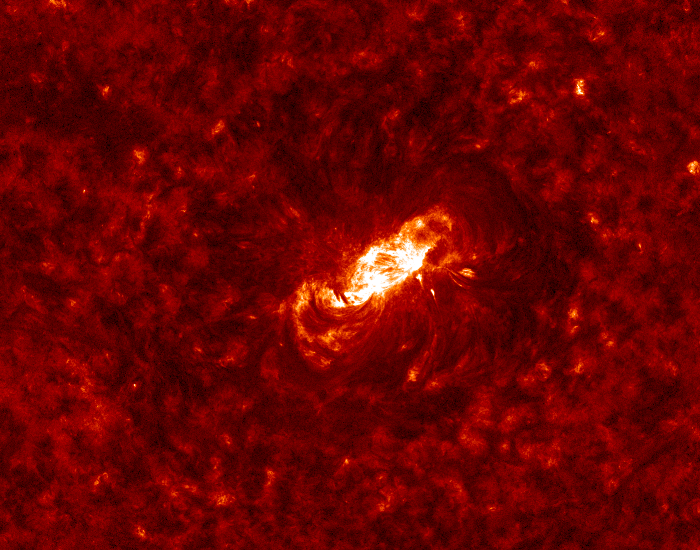}
\put(5.0,70){\color{white}{ \fontsize{8}{9}\selectfont 304 {\AA}}}
\end{overpic}
\begin{overpic}[scale=0.1,angle=0,width=4cm,keepaspectratio]{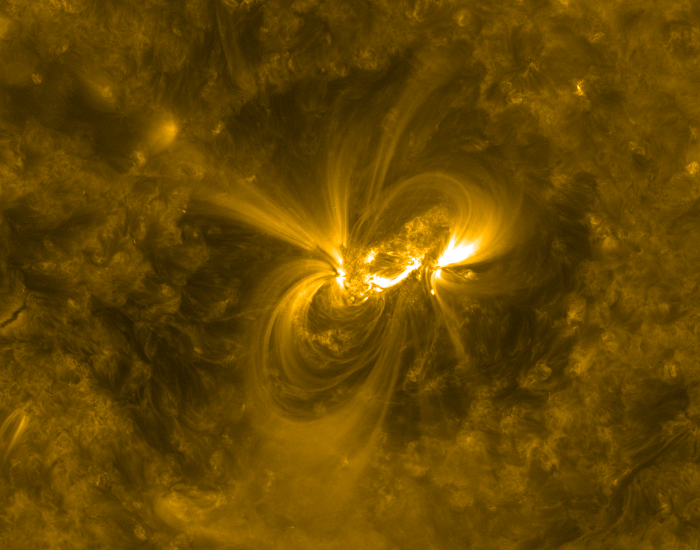}
\put(5.0,70){\color{white}{ \fontsize{8}{9}\selectfont 171 {\AA}}}
\end{overpic}
\begin{overpic}[scale=0.1,angle=0,width=4cm,keepaspectratio]{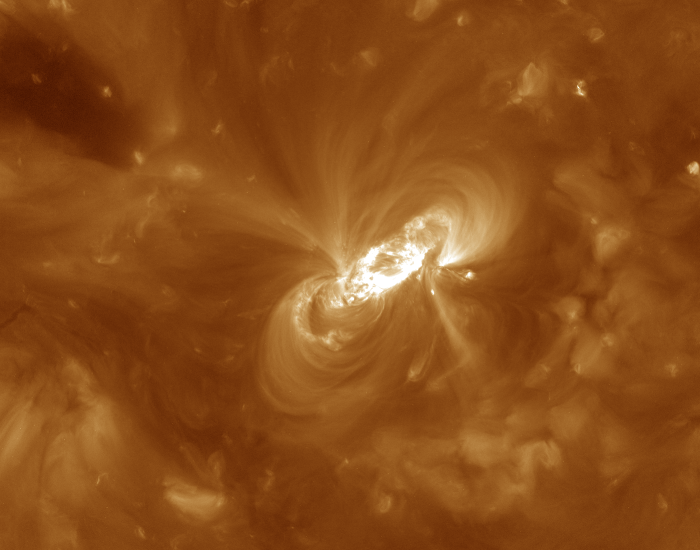}
\put(5.0,70){\color{white}{ \fontsize{8}{9}\selectfont 193 {\AA}}}
\end{overpic}
\begin{overpic}[scale=0.1,angle=0,width=4cm,keepaspectratio]{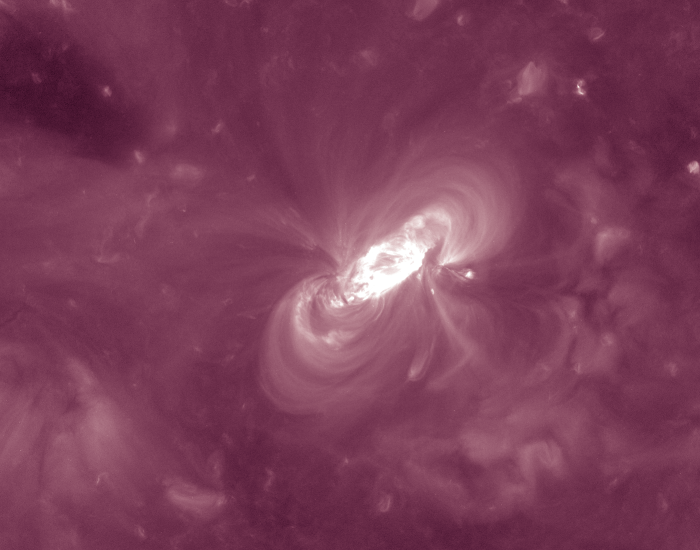}
\put(5.0,70){\color{white}{ \fontsize{8}{9}\selectfont 211 {\AA}}}
\end{overpic}\\
\begin{overpic}[scale=0.1,angle=0,width=4cm,keepaspectratio]{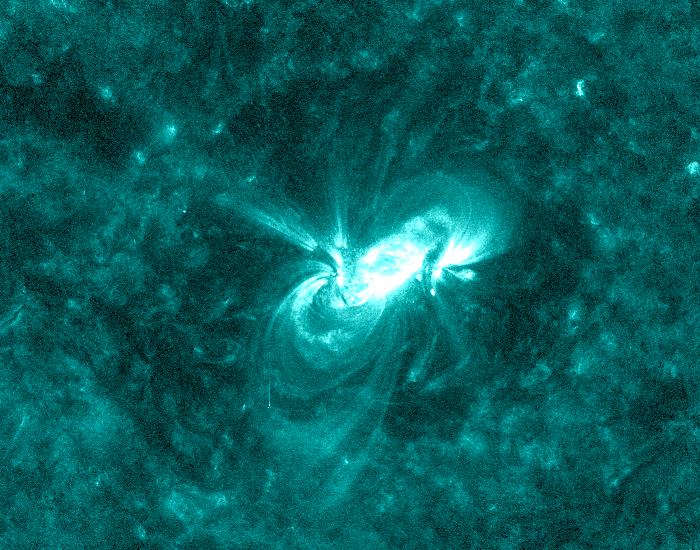}
\put(5.0,70){\color{white}{ \fontsize{8}{9}\selectfont 131 {\AA}}}
\end{overpic}
\begin{overpic}[scale=0.1,angle=0,width=4cm,keepaspectratio]{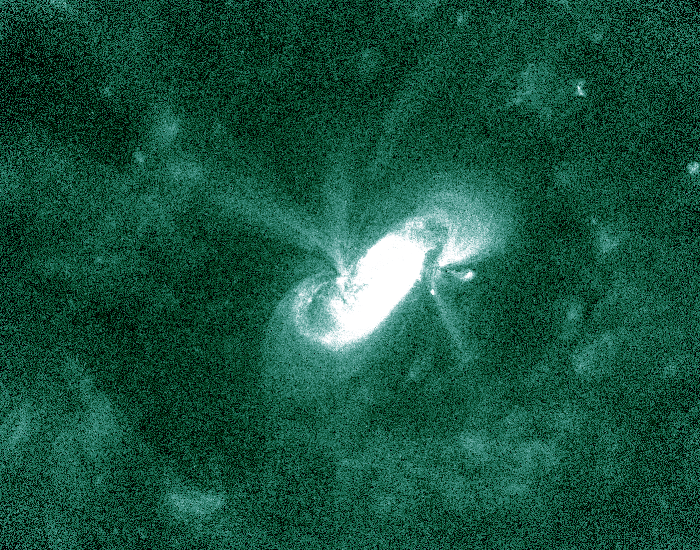}
\put(5.0,70){\color{white}{ \fontsize{8}{9}\selectfont 94 {\AA}}}
\end{overpic}
\begin{overpic}[scale=0.1,angle=0,width=4cm,keepaspectratio]{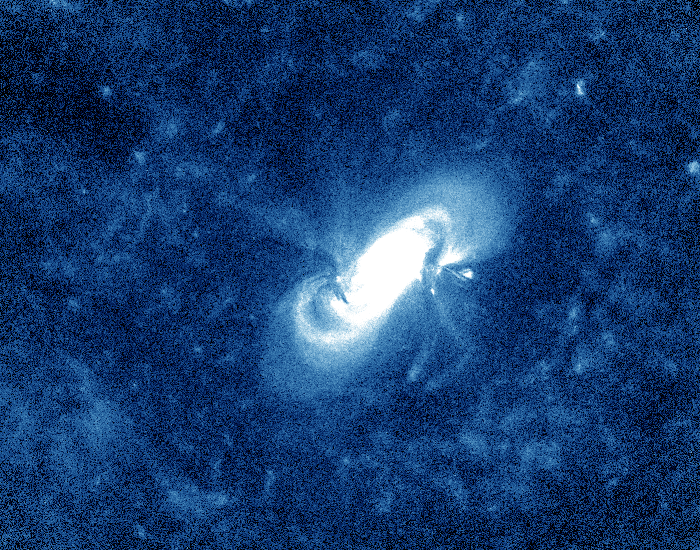}
\put(5.0,70){\color{white}{ \fontsize{8}{9}\selectfont 335 {\AA}}}
\end{overpic}
\begin{overpic}[scale=0.1,angle=0,width=4cm,keepaspectratio]{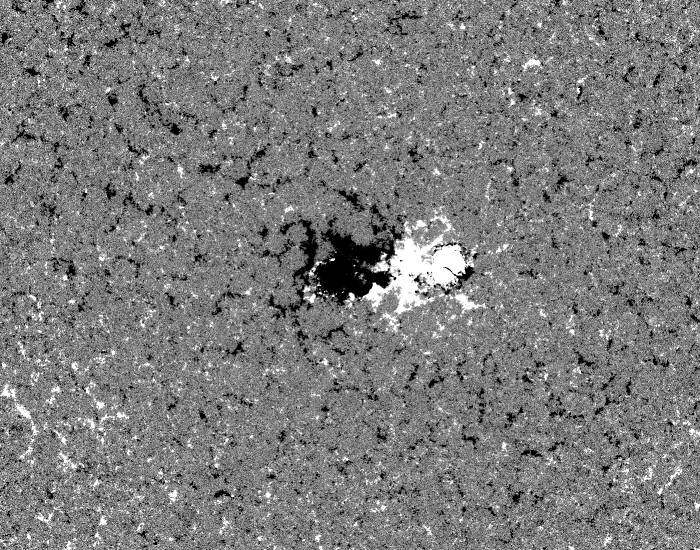}
\put(5.0,70){\color{white}{ \fontsize{8}{9}\selectfont HMI}}
\end{overpic}

\end{tabular}

\caption{The active region  of Fig.~1 (NOAA AR~12699), from 2018 Feb~11 06:04~UT (Table~1 case 1), 
in seven \sdo/AIA EUV channels and an HMI magnetogram.  The cutout size is 840$''$~$\times$~660$''$. 
The dark moat-like region around the AR is most obvious in the 171~{\AA} image, which has the strongest 
response to emissions over the temperature range \textup{0.6---1.1}~MK\@.}
\label{case1}
\end{figure}

To further examine the moat region of case 1, in Figure~\ref{case1} we show cutouts of 
size 840$''$ $\times$ 660$''$ of this region for EUV images in different wavelengths, and
for the HMI magnetogram. We find that the area of low emission measure is most clearly
seen in the 171~{\AA} wavelength, whose response function peaks in the temperature range
of \textup{0.6---1.1}~MK\@. In contrast, the dark moat is substantially less prominent in the 193,
211 and 335 {\AA} wavebands. We will continue to refer to such features as  ``moats" or
``dark moats" throughout the rest of the paper, whereas the surrounding brighter areas
are refereed to as ``bright regions.'' The magnetogram shows an area of mixed polarity in the dark moat.

Based on Figure~\ref{case1} the moat appears to be most prominent in 171~\AA\@. 
Although  less prominent, it is also detectable in 304~\AA, 131~\AA, 94~\AA, and 335~\AA\@.  It is still less prominent in 193~\AA, with the moat visible but not standing
out. Strikingly, in  211~\AA\ the dark moat is nearly absent.  The reduced or total lack
of visibility of the moat  in at least two wavelengths, 193~\AA\ and 211~\AA, along with
it being apparent in 304~\AA, are marked differences from the situation with coronal
holes.  This suggests that the field is not largely open, as in the case of true coronal
holes.  A different physical process from that operating in coronal holes must be
responsible for the moats being dark.

\begin{figure}[!htb]
\center
\begin{overpic}[scale=0.1,angle=0,width=9cm,keepaspectratio]{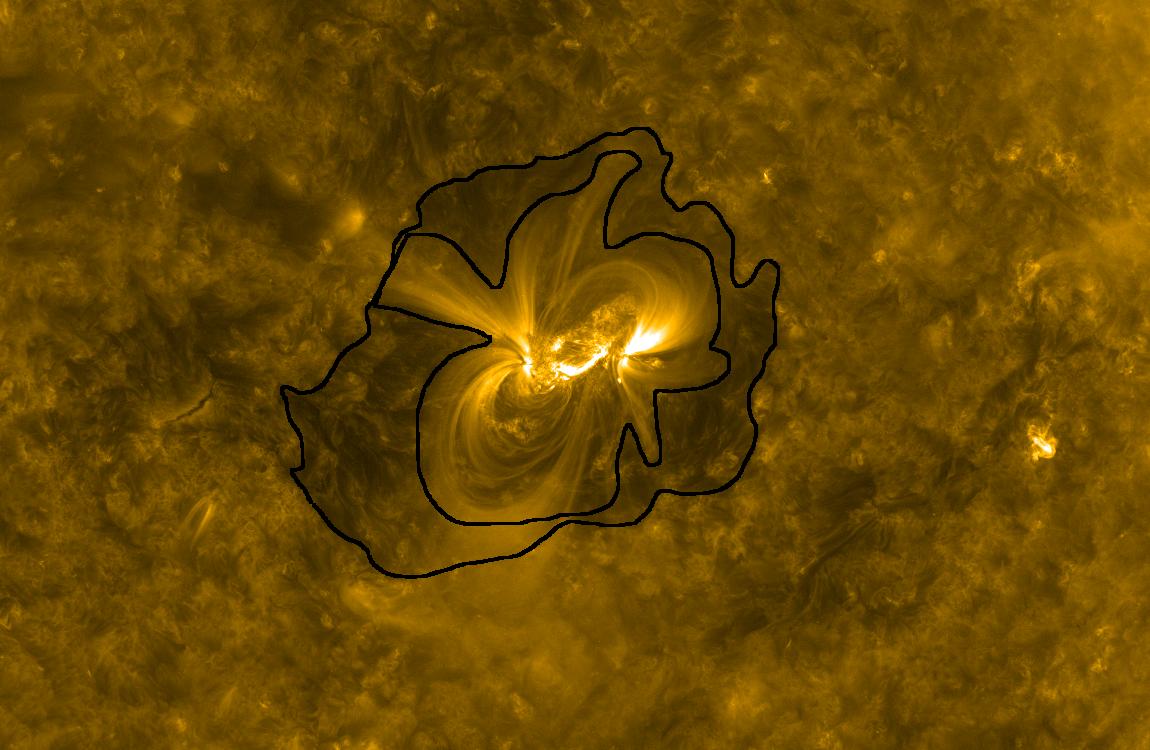}
\put(90.0,60){\color{white}{ \fontsize{8}{9}\selectfont 1}}
\end{overpic}
\caption{\textup{An \sdo/AIA 171~{\AA} image cutout from 2018 Feb 11 at 06:04~UT; case~1 of Table~1. The cutout size is 1380$''$~$\times$~900$''$. We have visually drawn a black boundary of the outer edge of the dark moat region. We have also drawn in black a contour showing our selection for the AR boundary (the inner boundary of the moat region), which surrounds the AR and the brightest 171~\AA\ loops extending out from 
the AR\@.}}
\label{moat_AR_bound}
\end{figure}

\begin{figure}[!htb]
%\vspace{11pc}
\center
\begin{tabular}{c c}  
\hspace{-0.6pc}
\begin{overpic}[scale=0.1,angle=0,width=6cm,keepaspectratio]{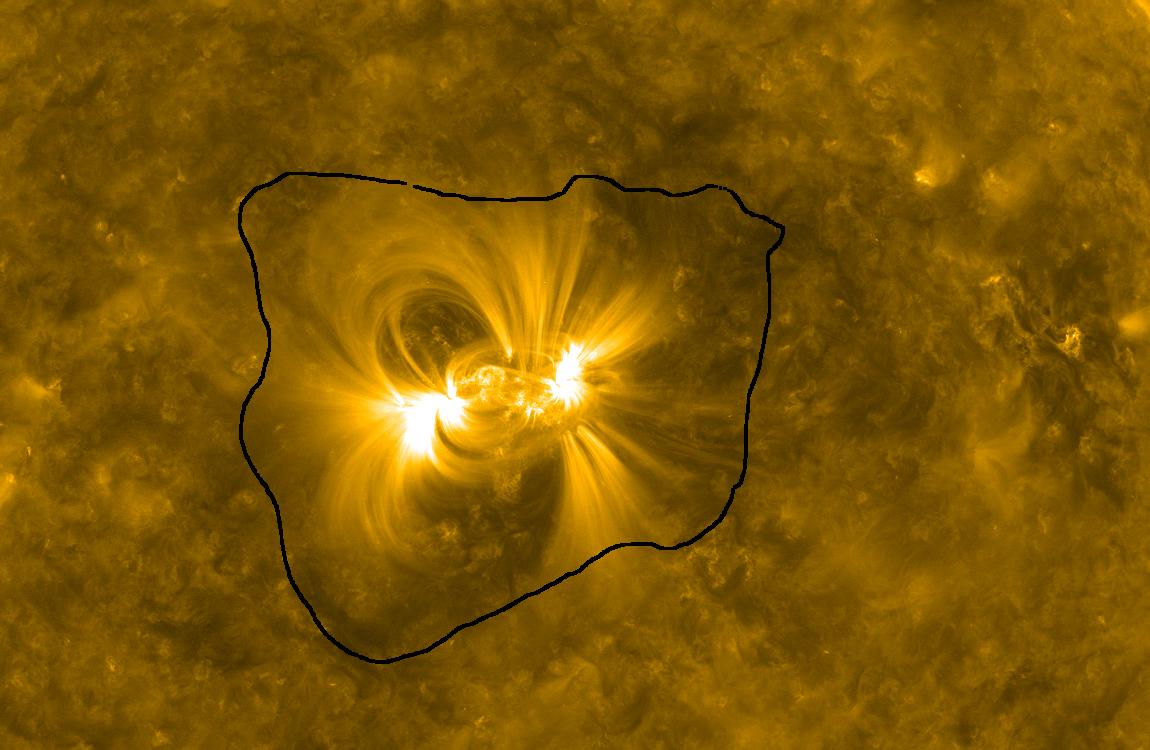}
\put(90.0,60){\color{white}{ \fontsize{8}{9}\selectfont 2}}
\end{overpic}
\begin{overpic}[scale=0.1,angle=0,width=6cm,keepaspectratio]{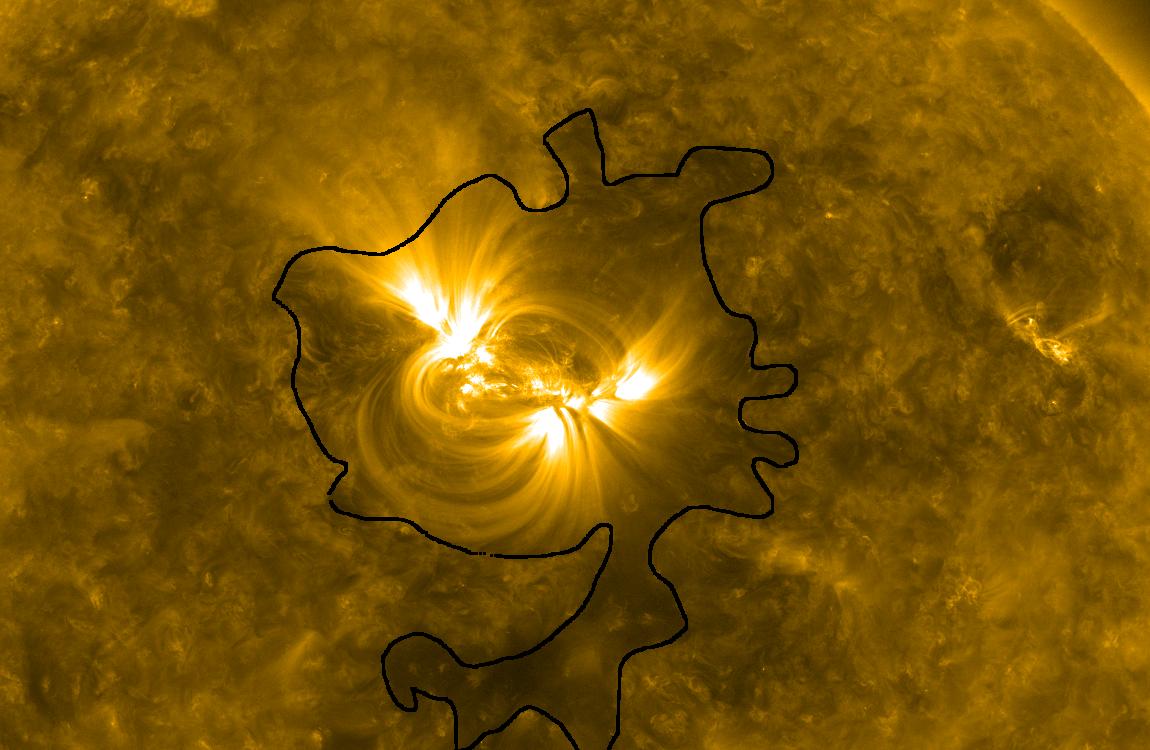}
\put(90,60){\color{white}{ \fontsize{8}{9}\selectfont 3}}
\end{overpic}\\
\hspace{-0.6pc}
\begin{overpic}[scale=0.1,angle=0,width=6cm,keepaspectratio]{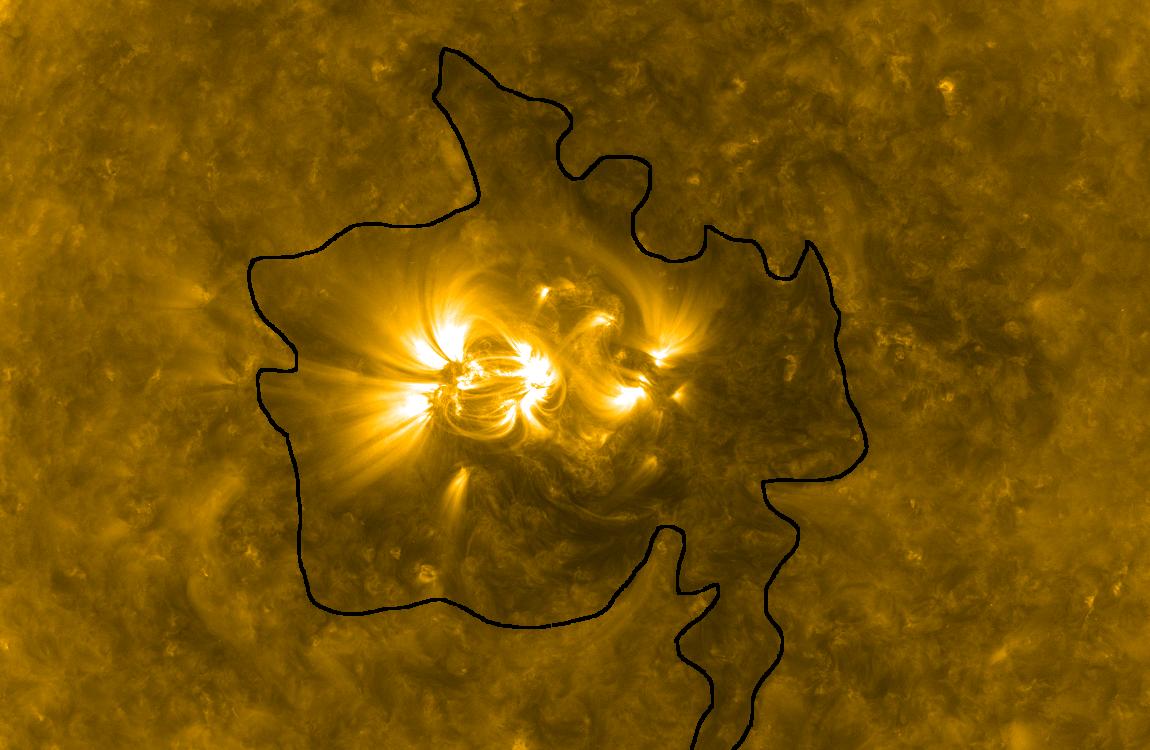}
\put(90.0,60){\color{white}{ \fontsize{8}{9}\selectfont 4}}
\end{overpic}
\begin{overpic}[scale=0.1,angle=0,width=6cm,keepaspectratio]{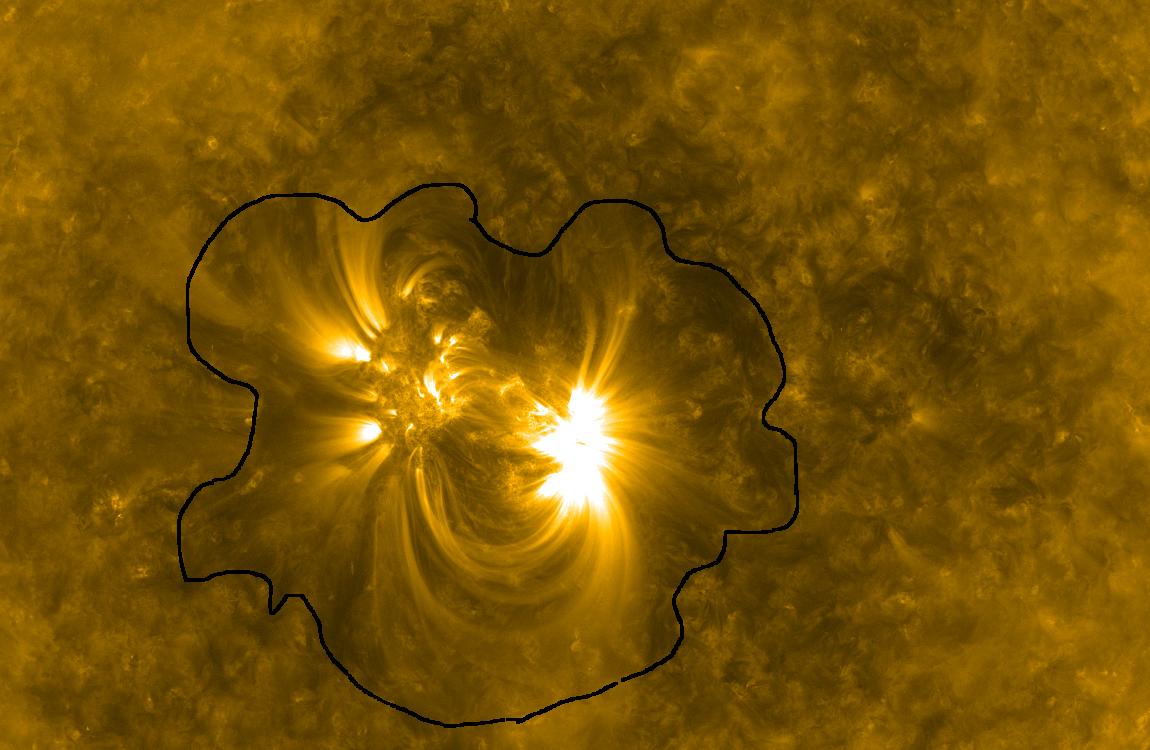}
\put(90.0,60){\color{white}{ \fontsize{8}{9}\selectfont 5}}
\end{overpic}\\

\begin{overpic}[scale=0.1,angle=0,width=6cm,keepaspectratio]{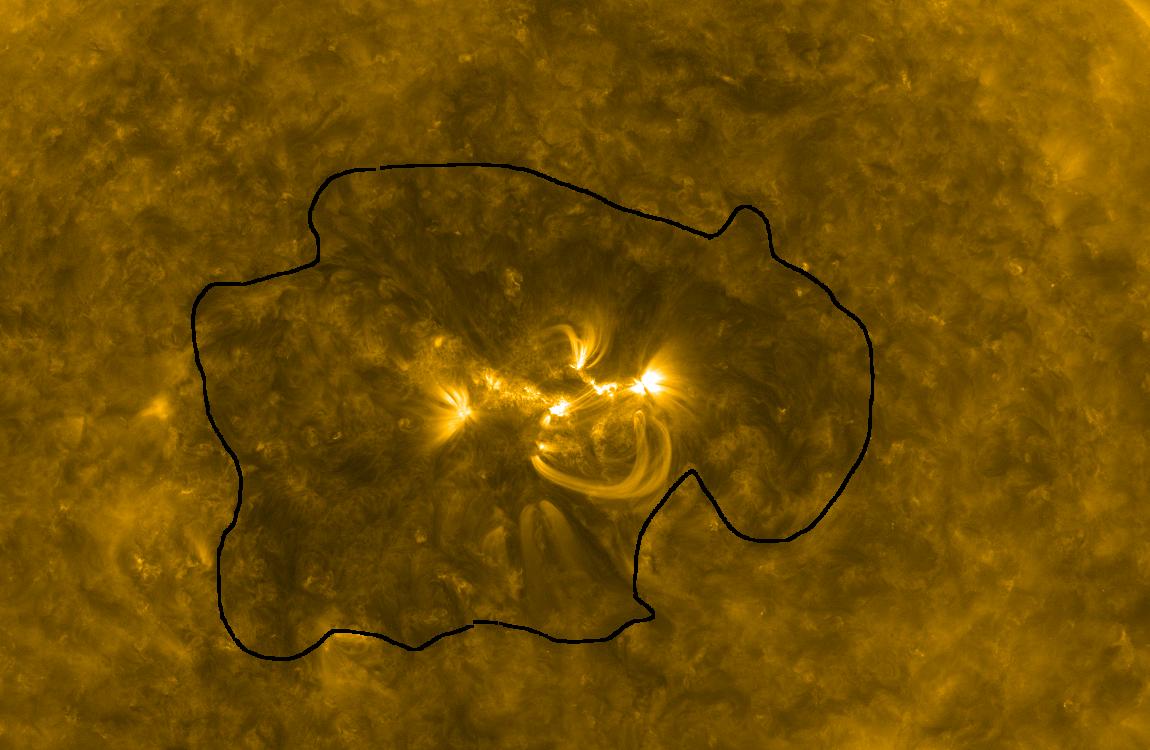}
\put(90.0,60){\color{white}{ \fontsize{8}{9}\selectfont 6}}
\end{overpic}
\begin{overpic}[scale=0.1,angle=0,width=6cm,keepaspectratio]{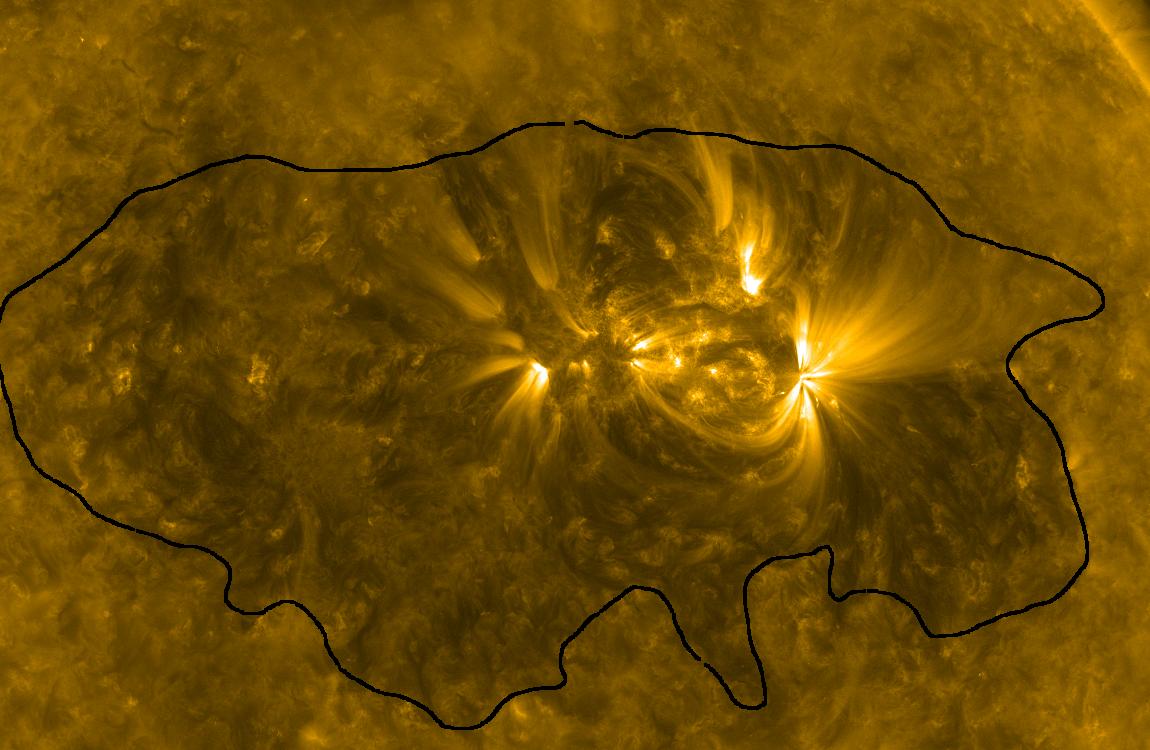}
\put(90.0,60){\color{white}{ \fontsize{8}{9}\selectfont 7}}
\end{overpic}
\end{tabular}

\caption{Similar to Fig.~\ref{case1}, but for Table~1 cases~2 to 7, and only showing AIA 171~{\AA} images.  The case number
for each appears in the panels, and the dates and times are those listed in Table~1 for the respective case.  Each cutout is 
1380$''$~$\times$~900$''$.  As with case~1 in Fig.~\ref{case1}, each of these cases also clearly show 
a low emission region around the AR\@. \textup{Here, we have visually drawn in black the boundary of the low emission areas.}}
\label{all171}
\end{figure}

\textup{To quantify the emission differences between moat and non-moat regions in different wavelengths, we find the average emissions in these regions for all 7 cases. To find these averages, we select the moat outer boundaries by eye and visually discard the area covered by the AR using 171 \AA\ images. The visually found moat and AR boundaries for Case~1 are shown in Figure~\ref{moat_AR_bound}. The outer moat boundaries for cases 2 to 7 can be seen in Figure~\ref{all171}. The external non-moat region is the surrounding area outside this manually selected moat region, and largely consists of quiet Sun. To reduce the effect of EUV limb-brightening in our average calculations, we only include the non-moat area within 60$^\circ$ of disk center. The results are presented in Table~2, where in each case the moat and AR-boundary contours determined from the 171~\AA\ 
image were applied to the corresponding images of the other channels. These values show that, on average, the moat regions are significantly darker in 171~\AA\ and 131~\AA, they are comparable in intensity to non-moat regions in 304~\AA, 95~\AA, and 335~\AA, and are brighter in 193~\AA\ and 211~\AA\ wavelengths. }

\begin{table}

\centering

\begin{tabular}{cccc}
\hline
Wavelength (\AA) &  Non-Moat (DN)     & Moat (DN)  & Relative Emission \\ \hline
304       & 8.1 $\pm$ 0.3   &  7.3 $\pm$ 0.6   & 0.90 $\pm$ 0.09 \\ \hline
171       & 212.6 $\pm$ 12.2 &  136.5 $\pm$ 23.1 & 0.64 $\pm$ 0.09 \\ \hline
193       & 138.9 $\pm$ 13.4 &  178.5 $\pm$ 25.4 & 1.28 $\pm$ 0.26\\ \hline
211       & 41.0 $\pm$ 6.1  &  67.4 $\pm$ 8.0  & 1.64 $\pm$ 0.29 \\ \hline
131       & 5.8 $\pm$ 0.1   &  4.4 $\pm$ 0.4   & 0.76 $\pm$ 0.07\\ \hline
94        & 0.9 $\pm$ 0.1   &  0.9 $\pm$ 0.1   & 0.96 $\pm$ 0.13\\ \hline
335       & 0.9 $\pm$ 0.1   &  1.0 $\pm$ 0.1   & 1.13 $\pm$ 0.15\\ \hline
\end{tabular}
\caption{Average emissions in different AIA wavelength measurements for all seven cases, along with the standard deviation across these cases. Relative Emission is the ratio: Moat Emission/Non-Moat (quiet Sun) Emission. Standard deviation of this ratio is found using the Taylor expression of the second moment.}
\label{table2}
\end{table}

In Figure~\ref{all171}, we show similar cutouts of size 1380$''$ $\times$ 900$''$ around ARs
for Table~1 cases~2 to~7 in the 171~\AA\ wavelength. Dark moat regions are present around
the ARs in each of the  cases. These moats vary greatly in shape and size for the
different ARs, and they often are  not symmetric. Moreover, the character of the apparent
boundary can vary in differing regions: case~1 (Fig.~\ref{case1}) has a relatively sharp
boundary between the  dark moat and the surrounding bright regions; in contrast, cases~2
through~6 show a more gradual transition from the dark moat to the bright regions, with
case~7 being closer to case~1 in this regard.

%\begin{comment}
%\end{comment}

%\begin{comment}
%\end{comment}

\subsection{DEM analysis}\label{DEM_section}

In Figure~\ref{case1}, the AR dark moat is more prominent in some of the AIA passbands than others.  This is a
reflection of the temperature distribution of the coronal plasma along the line-of-sight from
the \sdo\ satellite to the moat on the Sun, since the respective filters respond
preferentially to plasmas respectively radiating at differing temperatures. Although the
intensity in different AIA  passbands peaks at particular temperatures, their response
functions have contributions  from plasma from over a larger range of temperatures.  \textup{The response functions of different channels of AIA as a function of temperature can be seen in Fig.~1 of \citet{Cheung2015}}. We
can quantitatively examine  the 
distribution with temperature of the plasma along the line of sight from
the observer through any imaged solar feature, by estimating the differential emission measure
(DEM) function \citep{Withbroe78, Boerner2012}.  From \citet{Cheung2015}, the intensity in each pixel
from AIA wavelength-channel $i$ in an image can be written

\begin{equation}\label{eq_em}
I_i = \int_{0}^{\infty} K_i (T)\,\, \rm{DEM}(T)\,\,\rm{d}T.
\end{equation}

\noindent Here, $K_i$ contains the response function for filter $i$, and accounts for the  atomic processes responsible for the emission from the plasma detected by that
filter; this is a function of the temperature, $T$\@.   By using the intensities in
different AIA filters for a given location on the Sun, it is possible to estimate the DEM, the distribution with temperature of the square of the density of the emitting
plasma along the line of sight, through inversion of the integral equation,
eq.~\ref{eq_em}.   We have done this for each of the regions in Table~1, using the
procedure outlined in  \citet{Cheung2015}. Their method, available in IDL SolarSoft as
routine $aia\_sparse\_em\_init.pro$, uses AIA data in 94, 131, 171, 193, 211, and
335~{\AA} wavelengths to give DEM maps, showing the total emission measure (EM) in
different temperature ranges, where EM is DEM$(T)$ integrated over a specified temperature range. \textup{We have detailed the method used by \citet{Cheung2015} to calculate EM in Appendix~\ref{appendix_A}.}

Figure~\ref{DEM_1} shows EM maps for the AR of Table~1 case~1, with the different panels
showing the total EM contained within the specified temperature ranges. \textup{In our 
implementation of the $aia\_sparse\_em\_init.pro$ \textup{routine}, we have used 21 log~T bins of size 0.1, with the lower limit of 5.7.}  The cutout  size
is 1380$''$ $\times$ 900$''$, and is centered on the AR\@.  The dark moat visible in
Figure~\ref{case1} is most clearly seen in the log temperature range 5.75 to 6.05,
consistent  with the coronal emission being greatly suppressed in the dark moat in
this  temperature range.

Figure~\ref{DEM_all} displays similar EM maps for Table~1 cases~2 to~7 in this same
temperature range ($\log (T)=$[5.75,6.05]).  In each of these cases, the dark moat
shows substantial EM depletion, consistent with the visual appearance in Figure~\ref{all171}.

To further study the dependence of emission measure on temperature, we sample areas in
the dark moats and surrounding quiet regions and compare the average EMs in the two areas. \textup{For dark moats, we sample area inside a visually drawn boundary and visually exclude the area showing AR loops. The visually identified moat and AR boundaries for Case~1 were shown in Figure~\ref{moat_AR_bound}. For normal quiet-region areas, we consider all the area outside the moat boundary except the area outside $60^o$ from disk center to avoid sampling areas with limb-brightening in the EUV data.} 
The average EM
value is calculated in each of these two areas and plotted as  a function of temperature in the right
panel of Figure~\ref{DEM_B_vs_D}. The plot clearly shows less emission measure in
the dark moat compared to bright region over the  low-temperature range of 
$\sim$0.6---1.3~MK.

%\textup{The nominal errors returned by the code are negligible, because the EM values have been averaged over more than $10^5$ pixels.}

We have performed the same analysis for  all the other cases of Table~1 as well, and the
results are shown in Figure~\ref{DEM_B_vs_D_all}.  In each case, there is less EM
in the moat compared to the non-moat quiet region over roughly the same low-temperature range. 
\textup{There is a striking consistency in the difference in average EM values for all the seven cases. The moat has a lower EM value in the $\sim$0.7---0.9~MK range in all the cases. The moat regions also show EM values larger than the EM values in non-moat quiet regions in $\sim$1.3---5.0~MK range, as should be expected from the AR's hot coronal loops overlying the moat.}

\textup{To increase confidence in the EM features deduced with the $aia\_sparse\_em\_init.pro$ routine \citep{Cheung2015}, we also examined our Case~1 region using an independent DEM analysis method, namely that of \citet{Aschwanden13} and which is implemented in SolarSoft in the $aia\_teem\_map2.pro$ routine.
\citet{Aschwanden13} \citep[also see][]{Aschwanden11} take a more commonly used
``forward-fitting'' approach to solving the EM problem, whereby they assume a form for the 
EM as a function of log T (in the form of Gaussian or combinations of Gaussians in their 
case).  They then vary parameters of the assumed EM until the resulting predicted intensity best
matches the observed AIA intensity in each pixel, considering all six AIA coronal channels.  See 
\citet{Aschwanden11} and \citet{Aschwanden13} for details.}

\textup{A plot similar to the right panel of Figure~\ref{DEM_B_vs_D} from the \citet{Cheung2015} inversion method is shown in Figure~\ref{Mark_vs_Markus} from the \citet{Aschwanden13} inversion method. We see the same behavior of the difference in moat and non-moat EM curves in Figure~\ref{Mark_vs_Markus} as in the EM curves in Figure~\ref{DEM_B_vs_D}. The EM values calculated by these two methods do not match  perfectly. However, the peak EM values calculated by both methods match well,  though at slightly different temperatures. Thus, both the \citet{Cheung2015} method and the \citet{Aschwanden13} method independently show the moat regions to have reduced intensity over approximately 0.7---0.9 MK, and increased emission over approximately 1.3---5.0 MK, compared to the non-moat quiet regions.}

%\textup{In our analysis, we have limited the higher limit of the DEM analysis to 10.0~MK. The reason for that is that due to some undiagnosed reason, the calculated EM values (using both the methods described by \citet{Cheung2015}, and \citet{Aschwanden13}) start to rise again above this temperature, and with a peak around 25~MK. This is definitely a nonphysical phenomenon as there are no plasmas of substantial amounts in the corona at these temperatures, except in the case of solar flares.}
%This phenomenon is shown in the right panel of Figure~\ref{Mark_vs_Markus} for case 1 for non-moat region.}

%In some cases the drop is  more pronounced than in others, and the greatest contrast in
%emission over this temperature range (which roughly coincides with the lowest temperature
%window in the EM maps in Figs.~\ref{DEM_1} and~\ref{DEM_all}) ranges over
%$\sim$0.7---0.9~MK  (slightly lower than in the Fig.~\ref{DEM_B_vs_D} example).  In some cases, the dark moat has lower emission at higher temperatures too (case~3 being the
%most extreme example), but the intensity of the moats is consistently reduced in
%the low-temperatures window of $\sim$1~MK\@.

%\begin{comment}
\begin{figure}[!htb]
%\vspace{11pc}
\center
\begin{tabular}{c c c}  
\hspace{-0.6pc}
\begin{overpic}[scale=0.1,angle=0,width=5cm,keepaspectratio]{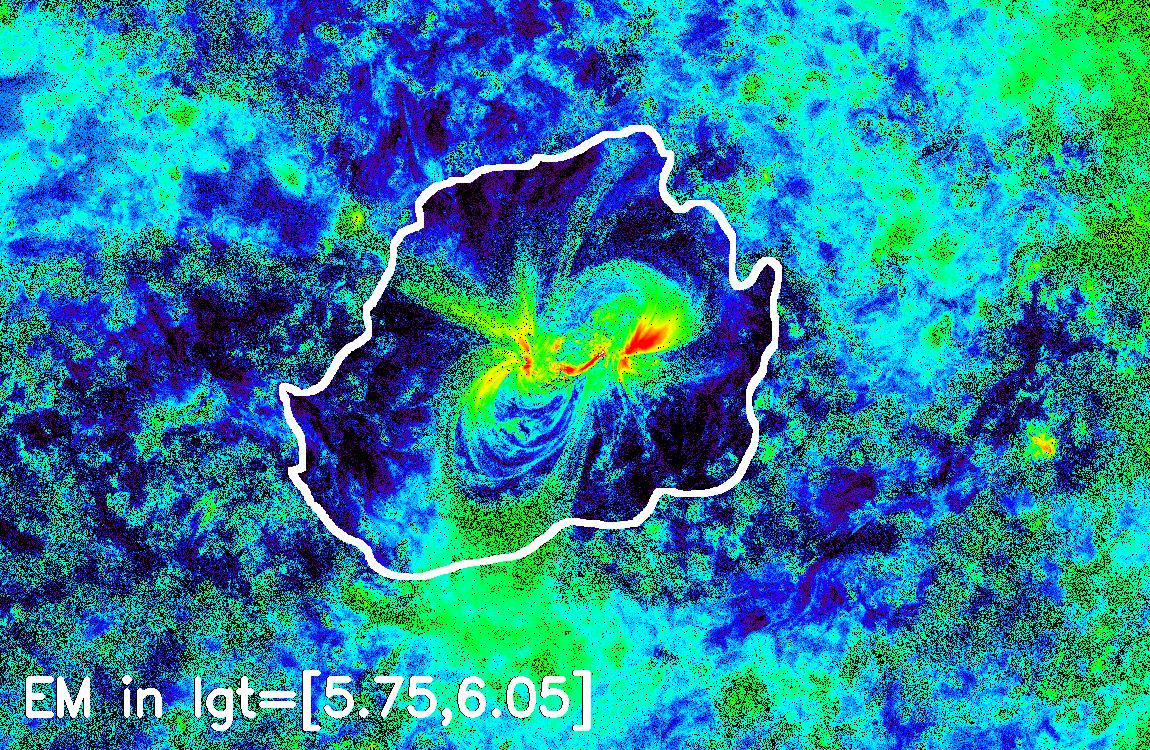}
\end{overpic}
\begin{overpic}[scale=0.1,angle=0,width=5cm,keepaspectratio]{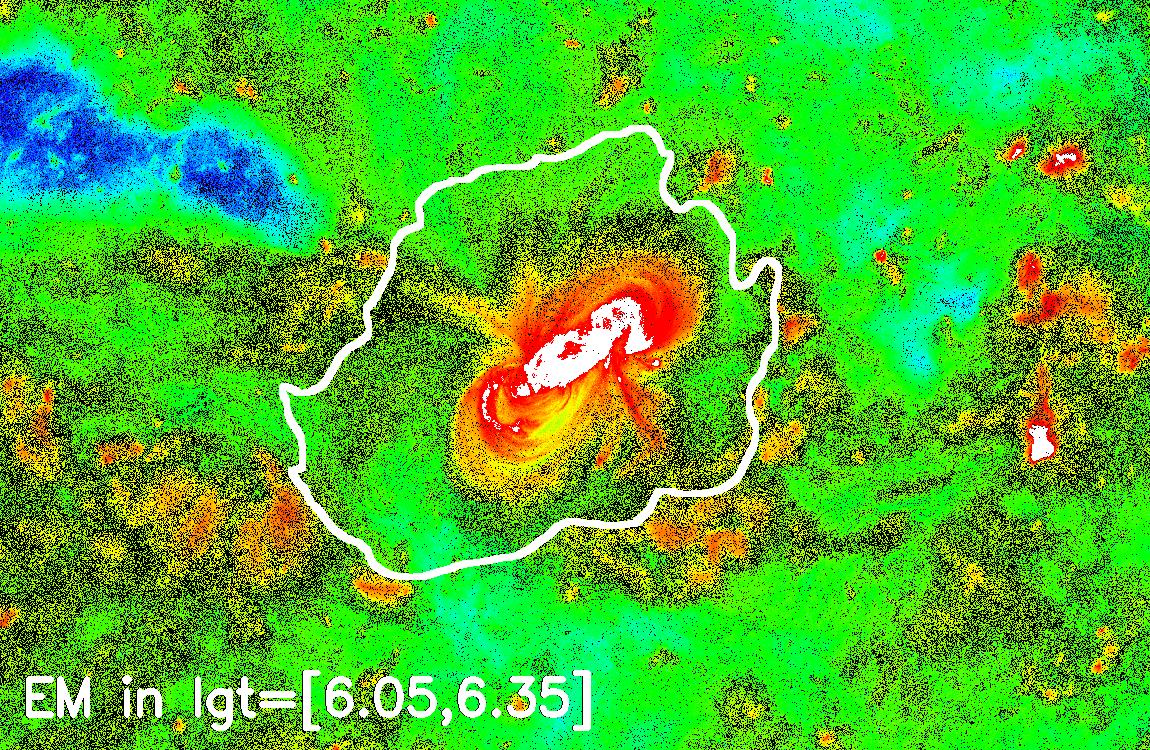}
\end{overpic}
\begin{overpic}[scale=0.1,angle=0,width=5cm,keepaspectratio]{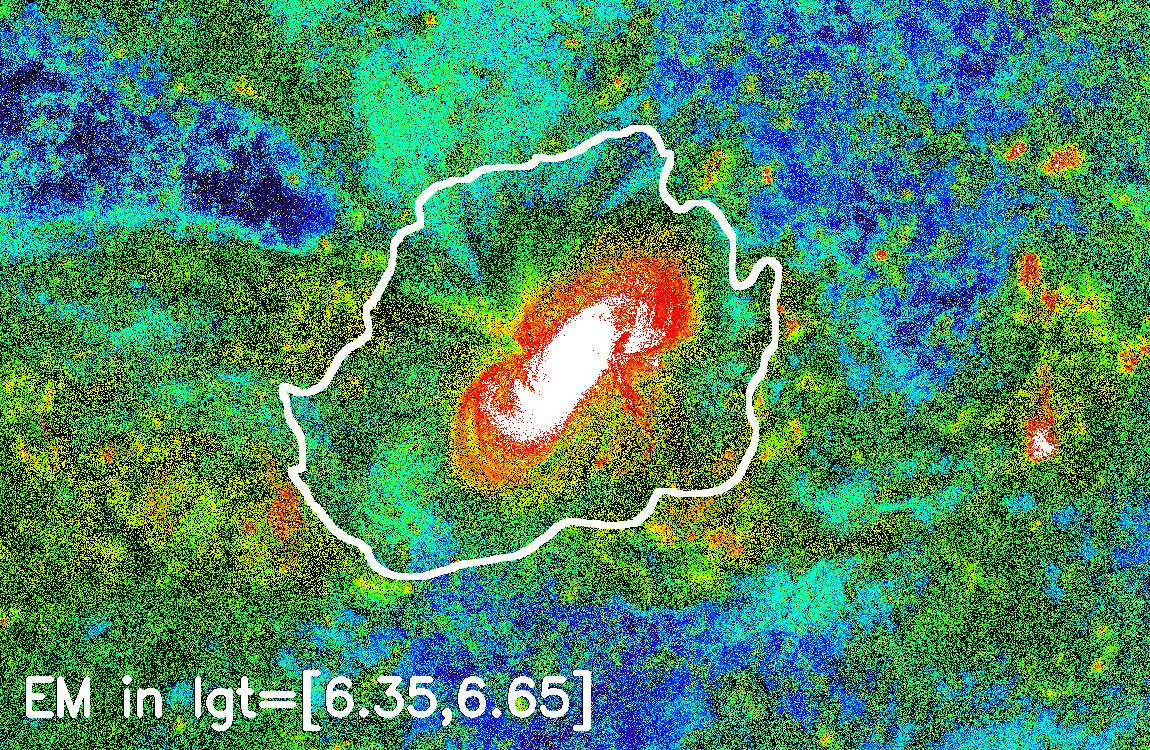}
\end{overpic}\\
\begin{overpic}[scale=0.1,angle=0,width=5cm,keepaspectratio]{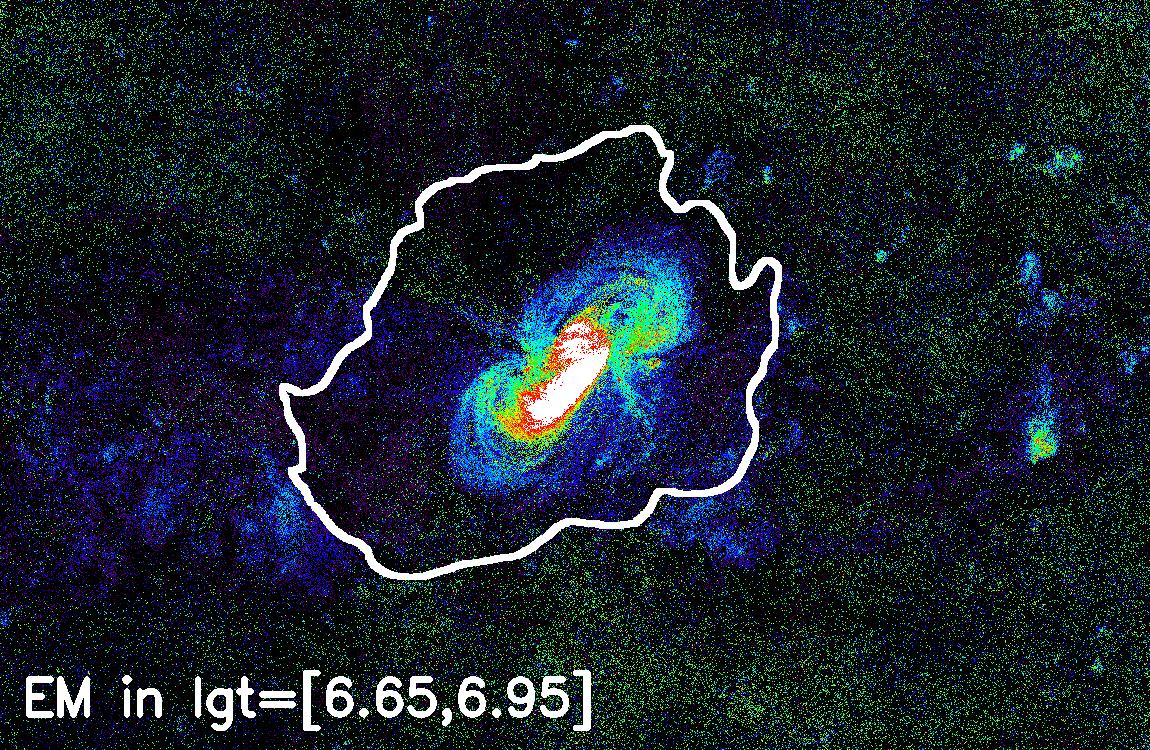}
\end{overpic}
\begin{overpic}[scale=0.1,angle=0,width=5cm,keepaspectratio]{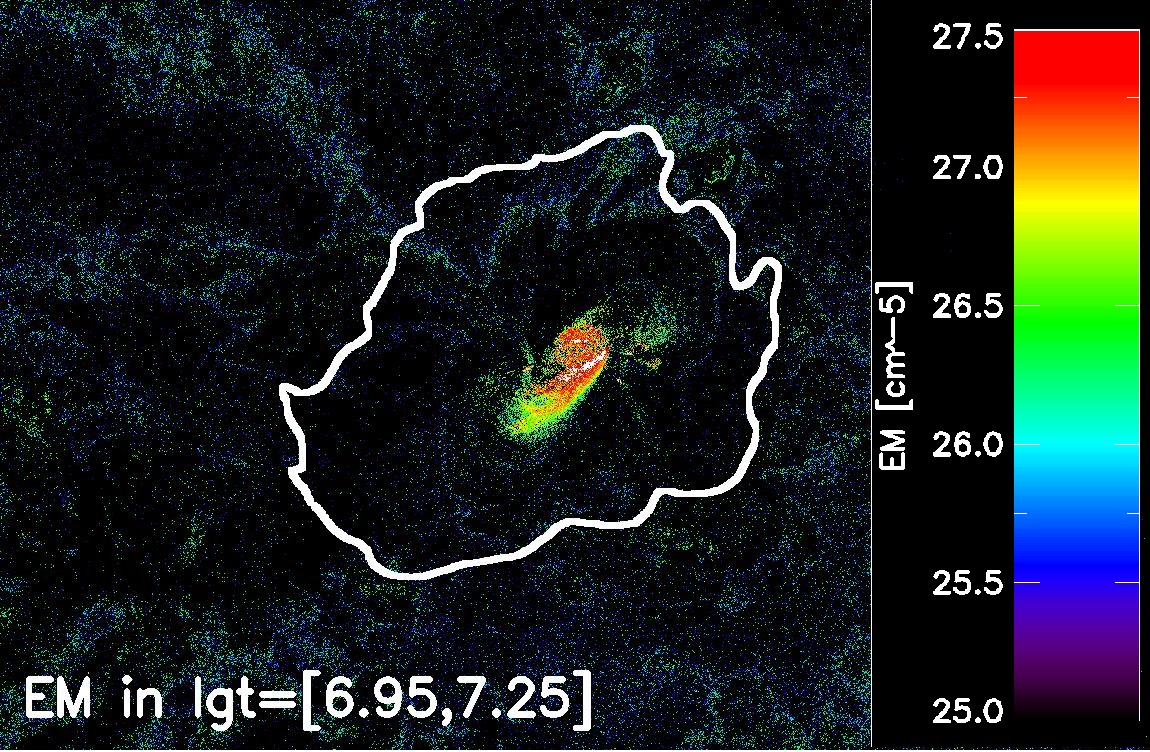}
\end{overpic}
\end{tabular}
\caption{Emission measure (EM) maps of AR 12699 (Table~1 case~1), made using AIA 94, 131, 171, 193, 211,
and 335~{\AA} images,  following the method described by \citet{Cheung2015}. All of the AIA images
used here are from 2018 Nov~02 06:04 UT\@.  The color-coding, indicated in the color bar,
represents the total emission measure (EM) contained within the  log~T range shown in the bottom left
corner of each panel. The cutout size is 1380$''$~$\times$~900$''$. From the color bar, white
represents the greatest emission; the low-emission dark moat is pronounced in the lowest temperature
range. \textup{We have visually drawn the boundary of the moat in the AIA 171 \AA\ image and overplotted it here as the white contour.}}
\label{DEM_1}

\end{figure}
%\end{comment}

%\begin{comment}
\begin{figure}[!htb]
%\vspace{11pc}
\center
\begin{tabular}{c c} 
\hspace{-0.6pc}
\begin{overpic}[scale=0.1,angle=0,width=6cm,keepaspectratio]{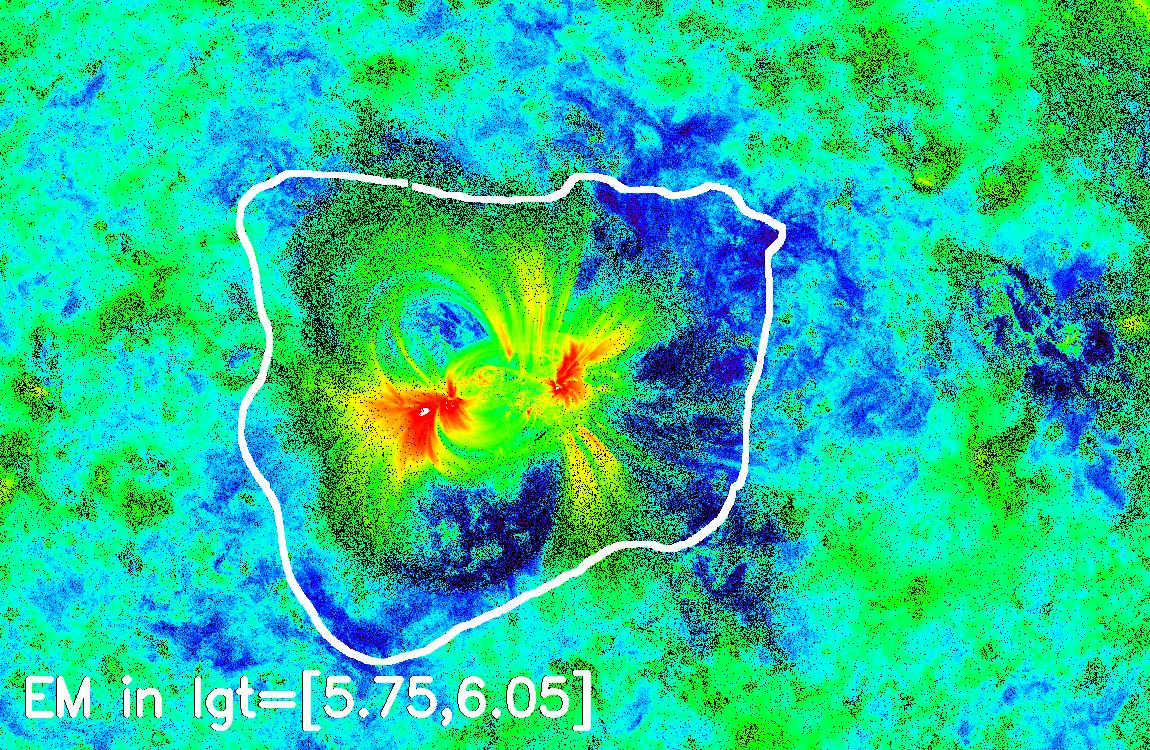}
\put(90.0,60){\color{white}{ \fontsize{8}{9}\selectfont 2}}
\end{overpic}
\begin{overpic}[scale=0.1,angle=0,width=6cm,keepaspectratio]{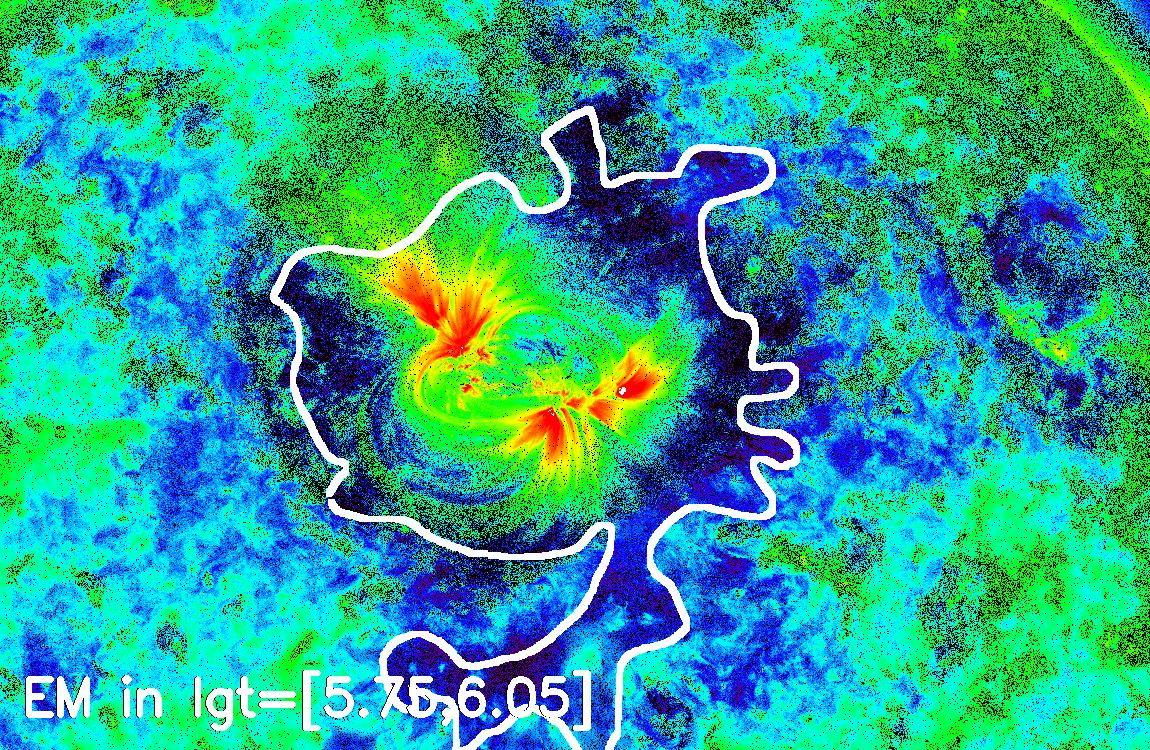}
\put(90,60){\color{white}{ \fontsize{8}{9}\selectfont 3}}
\end{overpic}\\
\hspace{-0.6pc}
\begin{overpic}[scale=0.1,angle=0,width=6cm,keepaspectratio]{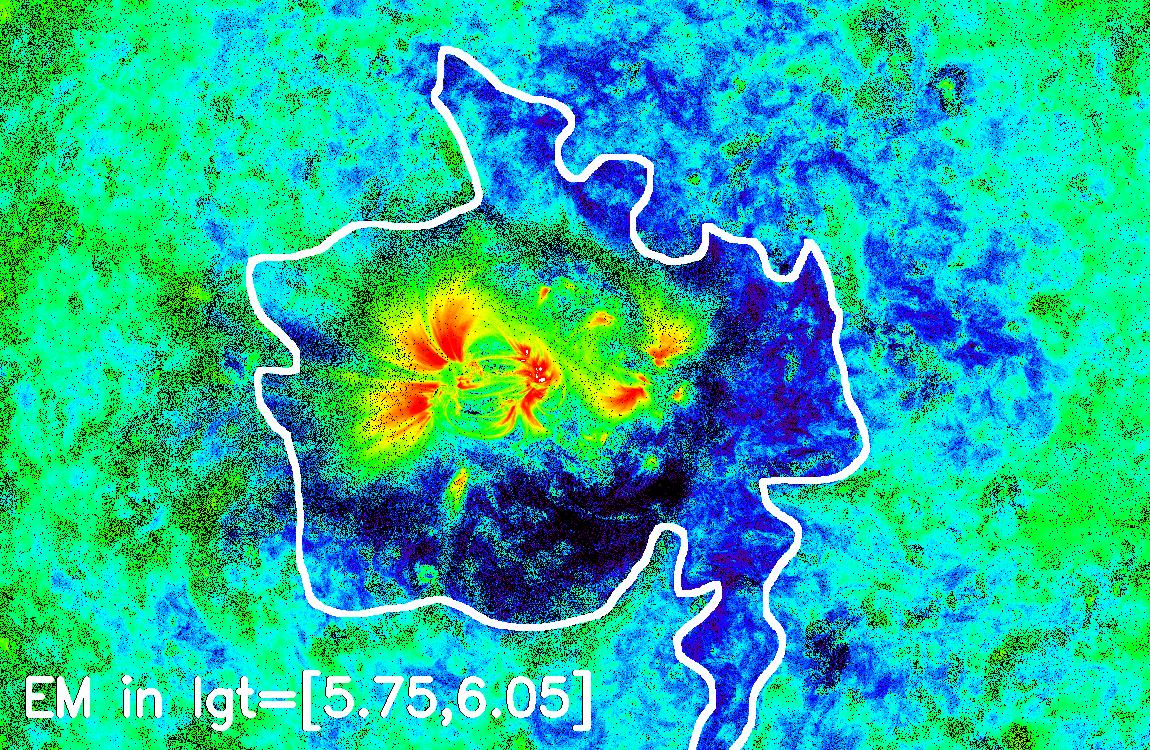}
\put(90.0,60){\color{white}{ \fontsize{8}{9}\selectfont 4}}
\end{overpic}
\begin{overpic}[scale=0.1,angle=0,width=6cm,keepaspectratio]{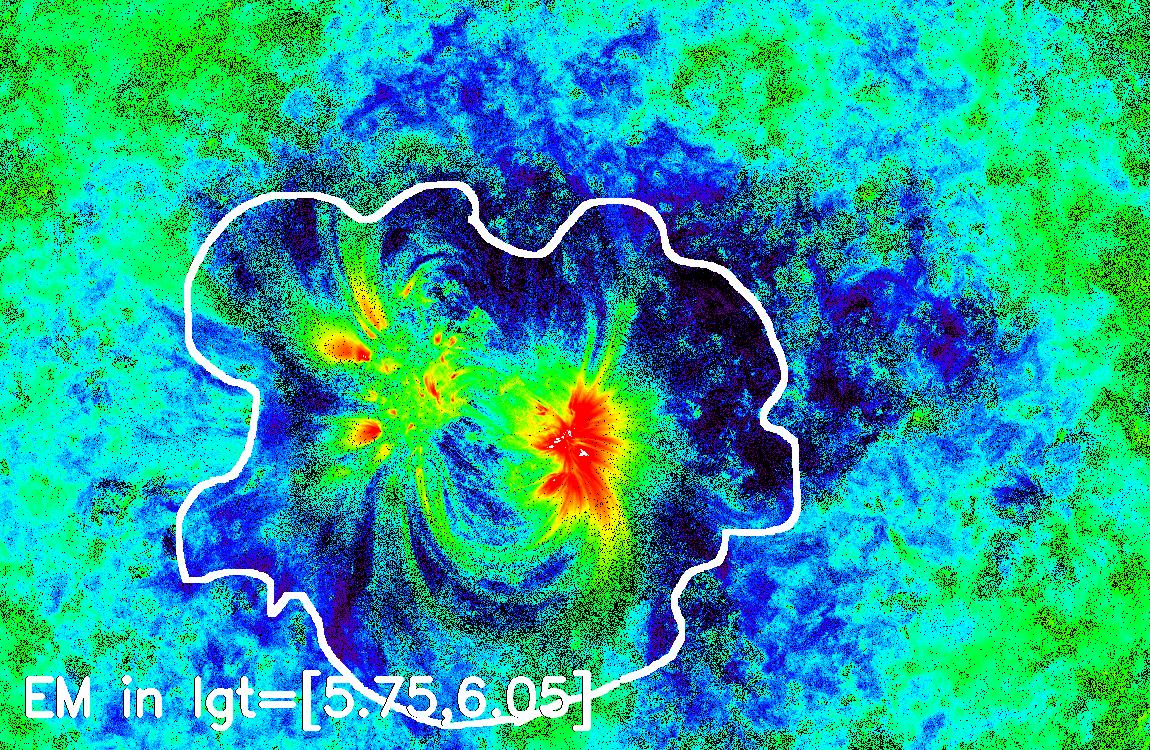}
\put(90.0,60){\color{white}{ \fontsize{8}{9}\selectfont 5}}
\end{overpic}\\

\begin{overpic}[scale=0.1,angle=0,width=6cm,keepaspectratio]{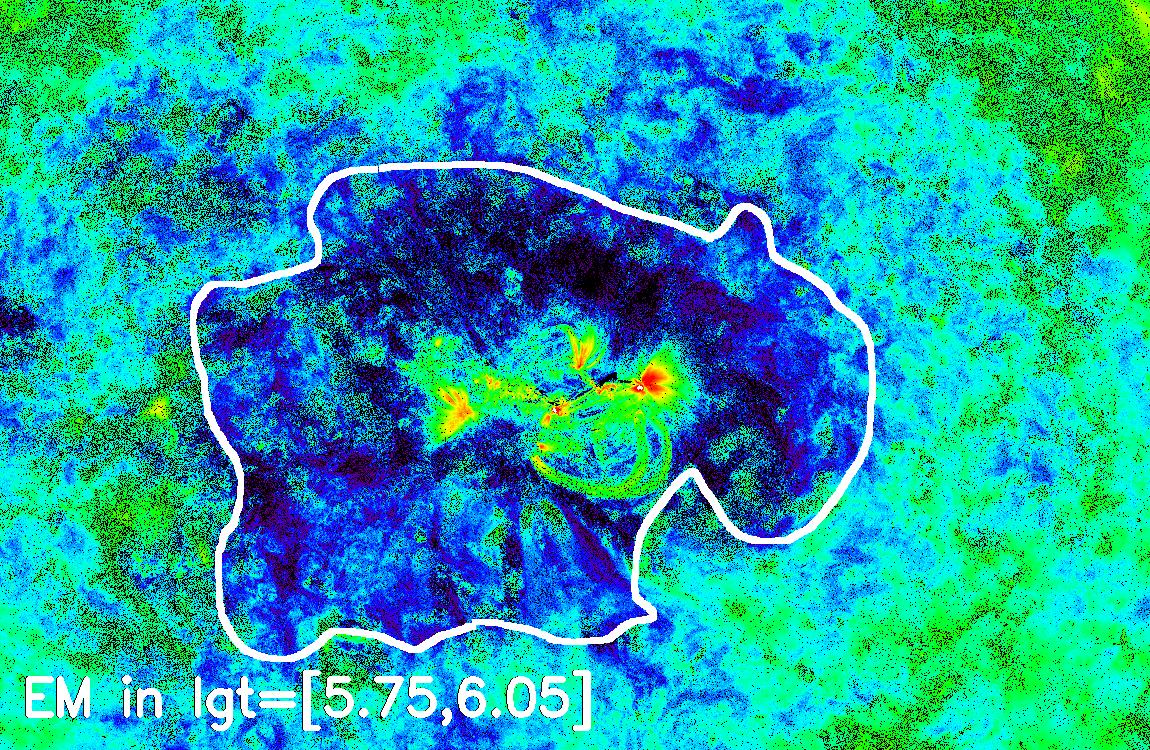}
\put(90.0,60){\color{white}{ \fontsize{8}{9}\selectfont 6}}
\end{overpic}
\begin{overpic}[scale=0.1,angle=0,width=6cm,keepaspectratio]{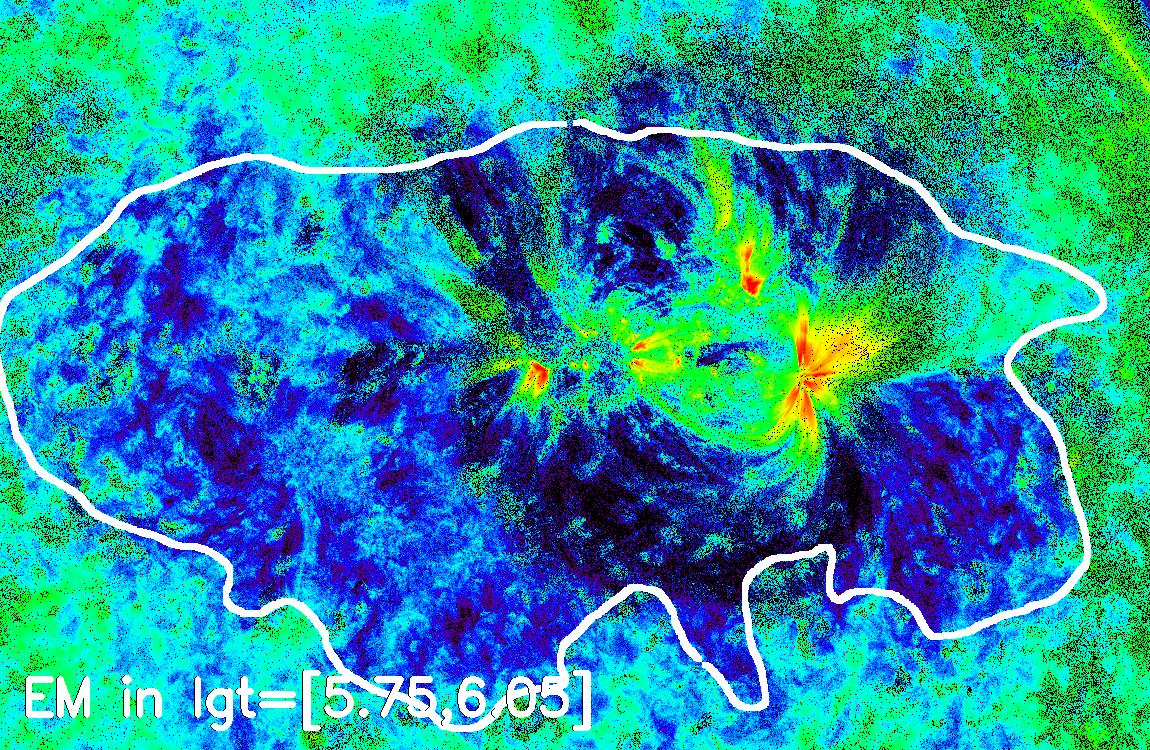}
\put(90.0,60){\color{white}{ \fontsize{8}{9}\selectfont 7}}
\end{overpic}
\end{tabular}

\caption{Similar to Fig.~\ref{DEM_1}, but showing EM maps for Table~1 cases 2 to 7,
showing EM in the log~T  range of 5.75 to 6.05, where the case numbers are indicated in
the panels.   The cutout size is 1380$''$~$\times$~900$''$  and the color-bar is the same as
in Fig.~\ref{DEM_1}.  Each case clearly shows low emission measure in the moat  that
appeared dark in the corresponding image in Fig.~\ref{all171}. This demonstrates that the
moat-like dark regions around ARs are dark because they are deficient in plasma over this temperatures range. \textup{We have visually drawn the boundary of the moat in each AIA 171 \AA\ image in Figure~\ref{all171} and overplotted it here as the white contour.}}
\label{DEM_all}
\end{figure}
%\end{comment}

%\begin{comment}
\begin{figure}[!htb]
%\vspace{11pc}
\center
\begin{tabular}{c c}  

\includegraphics[scale=0.1,angle=0,height=6cm,keepaspectratio]{DEM0_shrink.jpg} 
%\vspace{11pc}
\includegraphics[scale=0.1,angle=0,height=6cm,keepaspectratio]{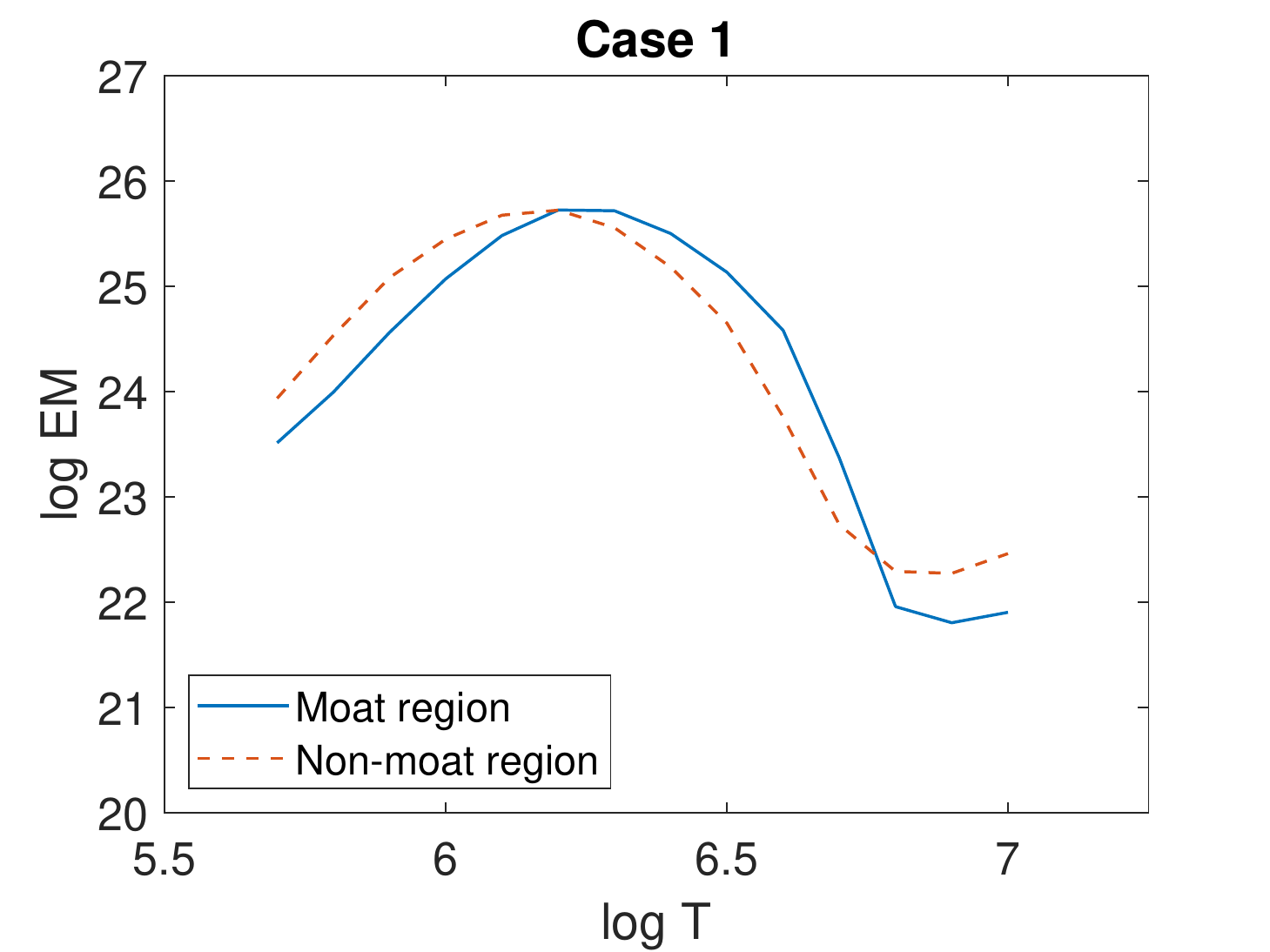}
\end{tabular}

\caption{(\textit{Left panel}) EM map of AR~12699, on 2018~Nov~02 at 06:04~UT (Table~1
Case 1), showing EM over  the log~T range of 5.75 to 6.05. The moat's outer boundary was visually drawn in the AIA 171 \AA\ image and is plotted here in white color. (\textit{Right panel}) Average  log~EM curves calculated
over the dark-region inside the moat boundary but excluding the AR (solid line) and outside the moat boundary but within 60 heliocentric degrees from the disk center (dotted line).  There is a definite depletion in the EM in the moat compared to that in the surrounding quiet region in the  low-temperature range. EM has the units of cm$^{-5}$ and T has units of Kelvin.}
\label{DEM_B_vs_D}
\end{figure}
%\end{comment}

%\begin{comment}
\begin{figure}[!htb]
%\vspace{11pc}
\center
\begin{tabular}{c c}  

\includegraphics[scale=0.1,angle=0,width=6cm,keepaspectratio]{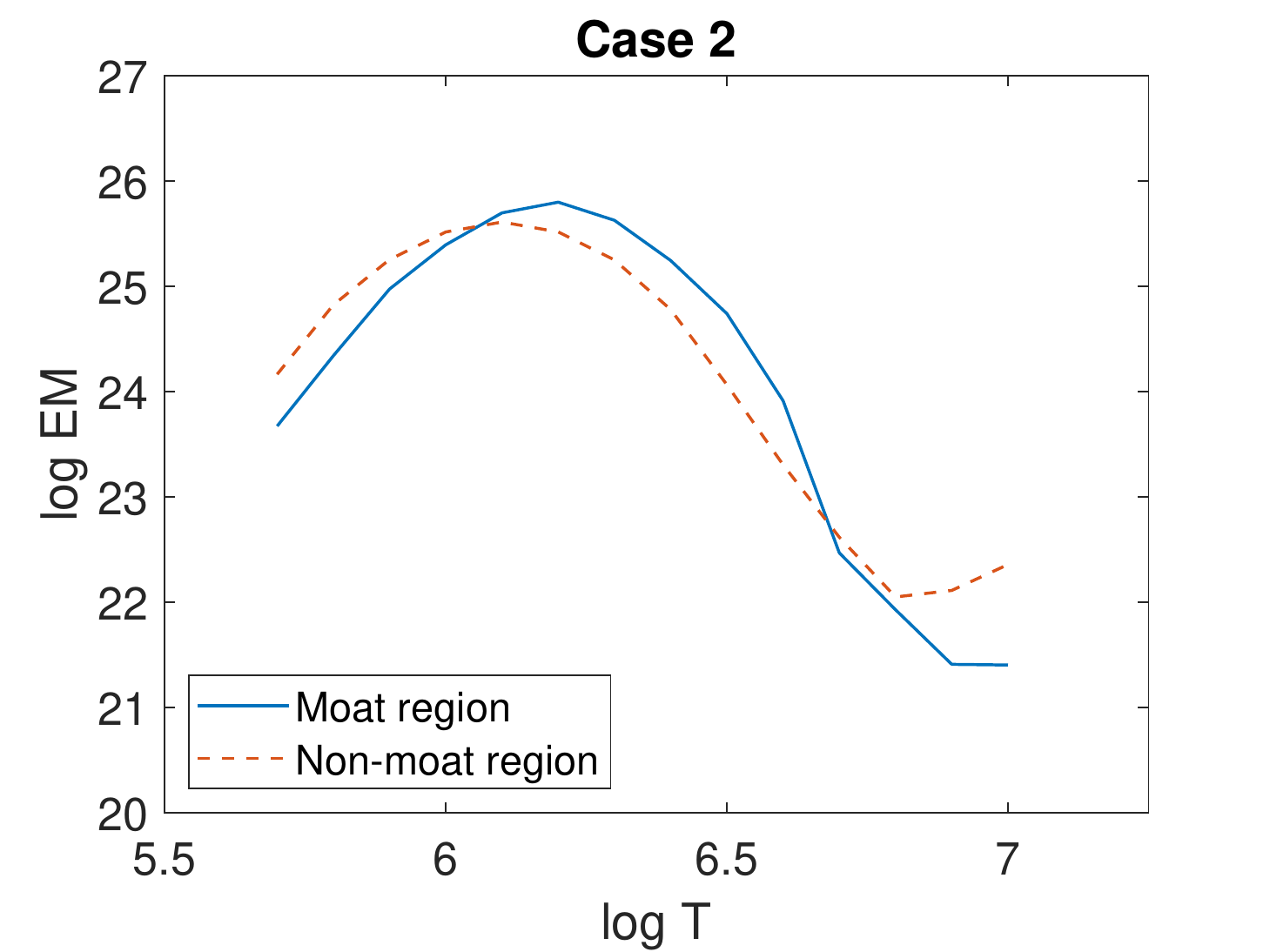} 
\includegraphics[scale=0.1,angle=0,width=6cm,keepaspectratio]{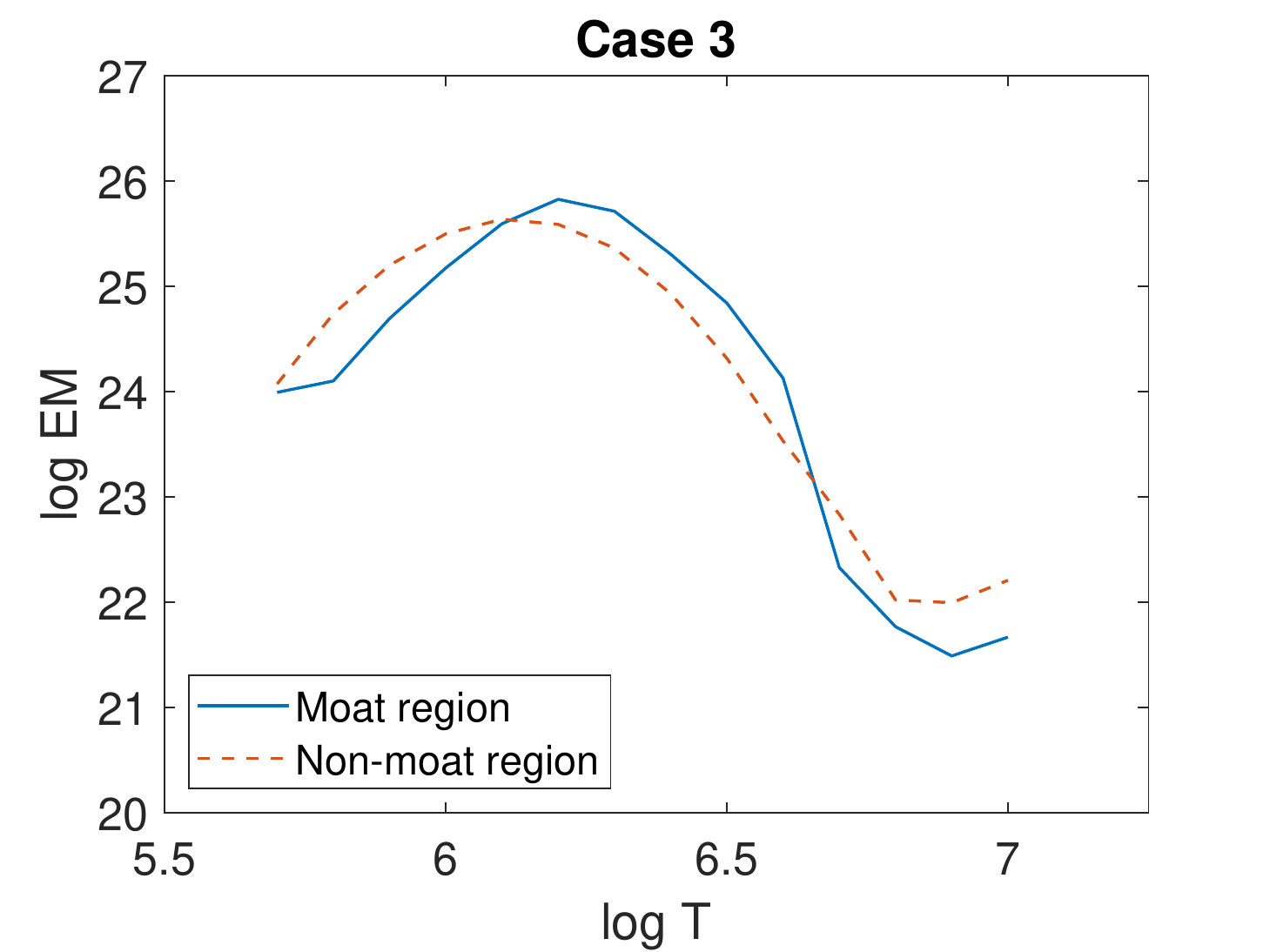} \\
\includegraphics[scale=0.1,angle=0,width=6cm,keepaspectratio]{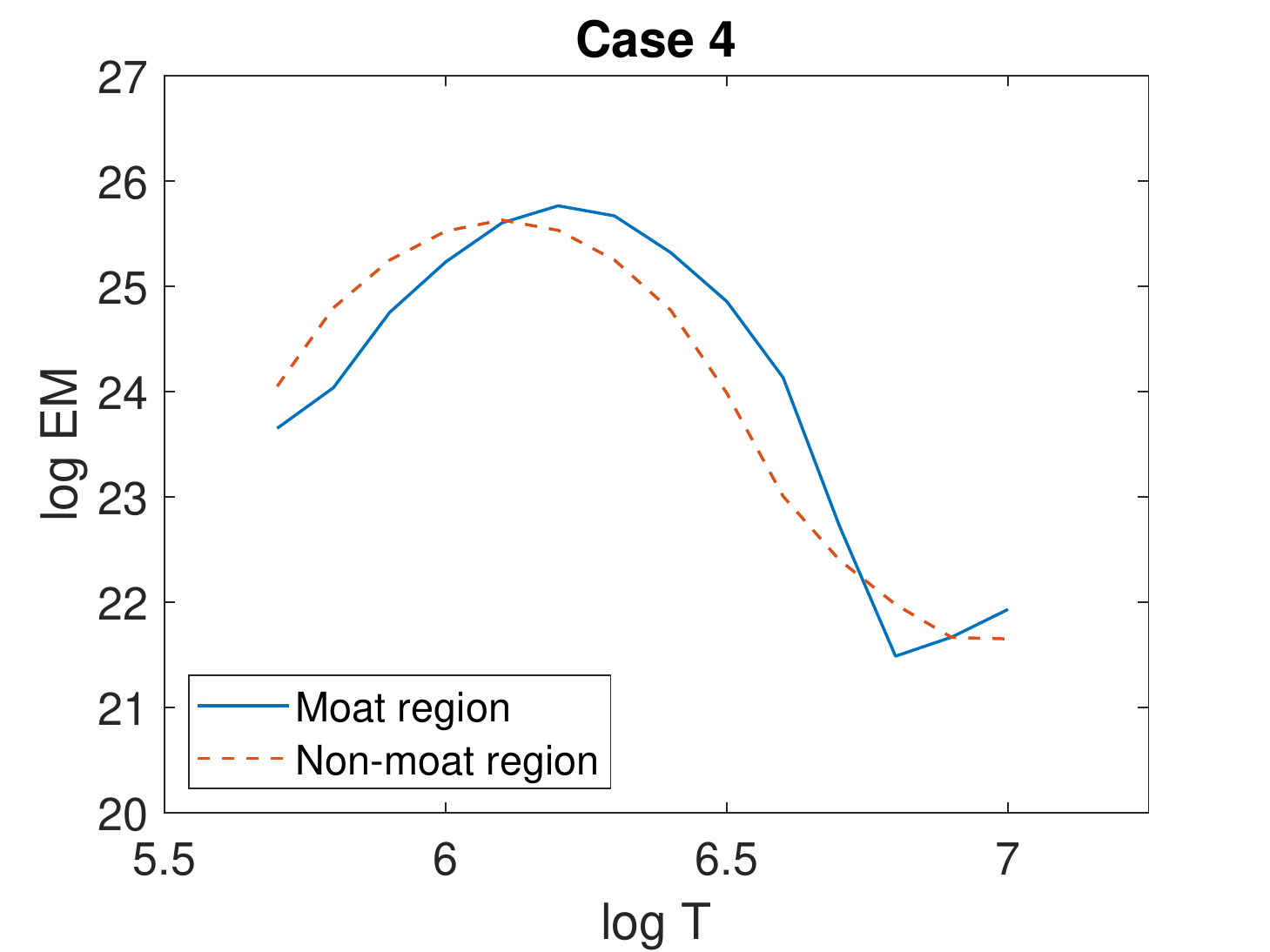}
\includegraphics[scale=0.1,angle=0,width=6cm,keepaspectratio]{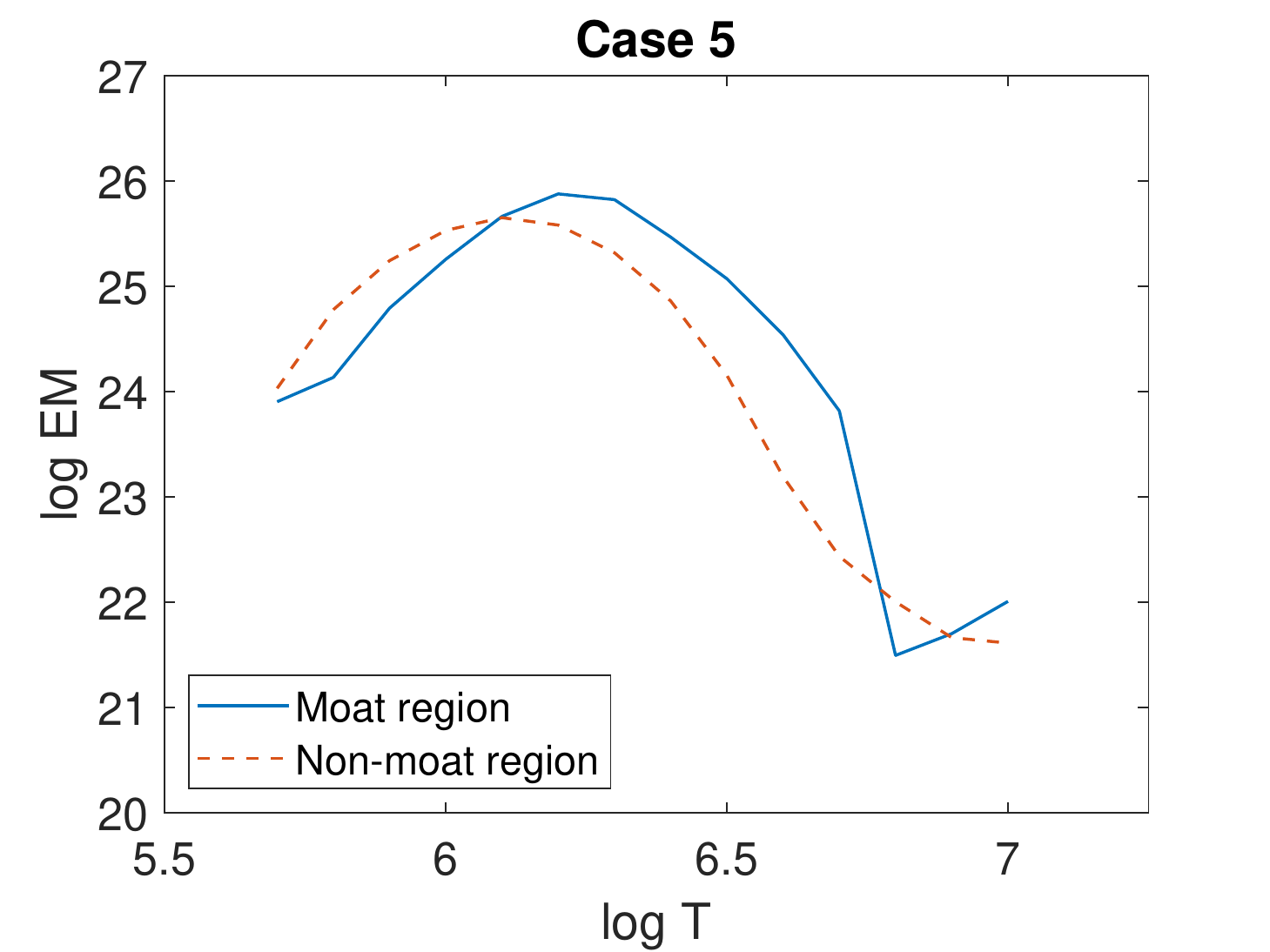} \\
\includegraphics[scale=0.1,angle=0,width=6cm,keepaspectratio]{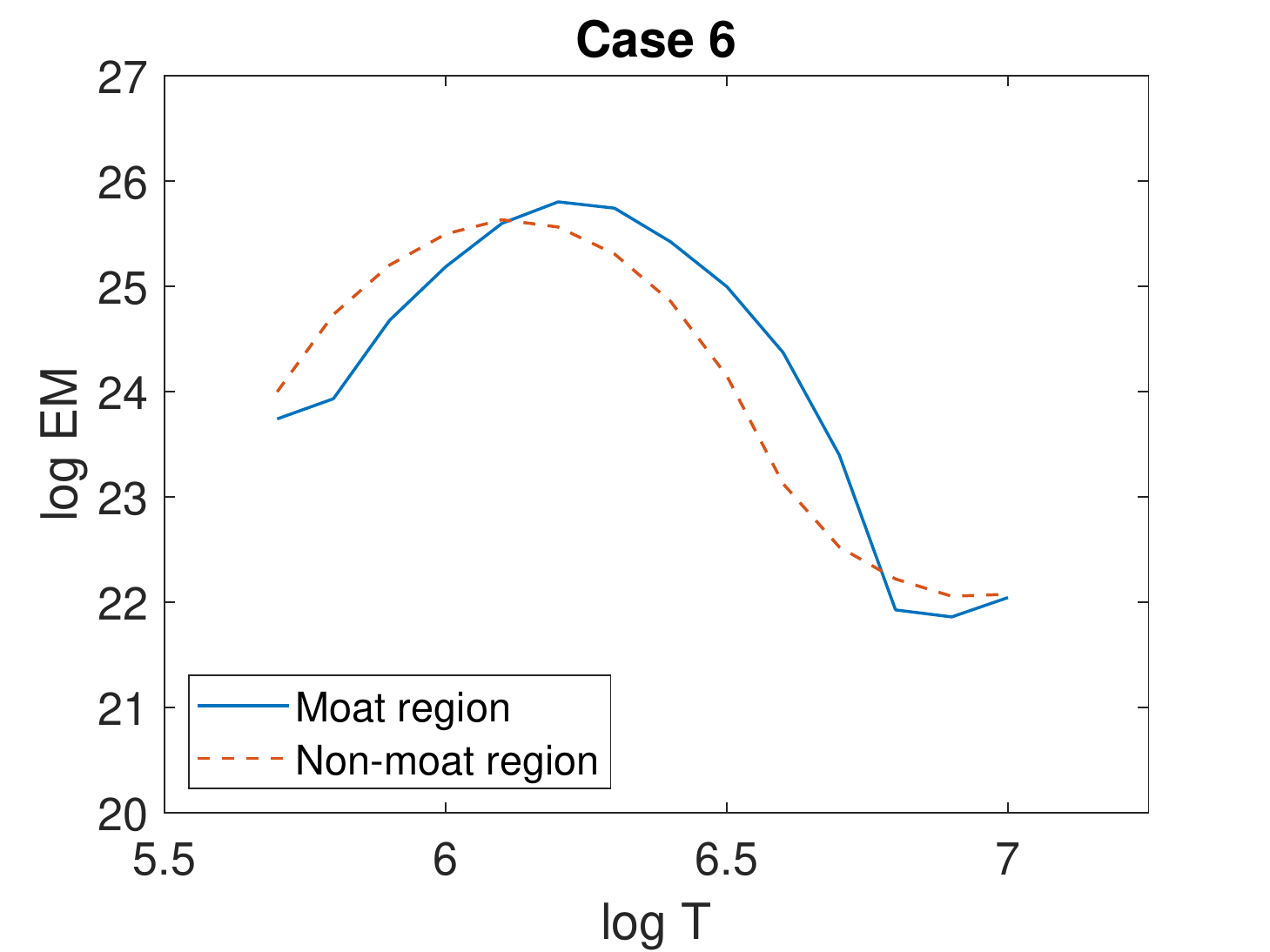}
\includegraphics[scale=0.1,angle=0,width=6cm,keepaspectratio]{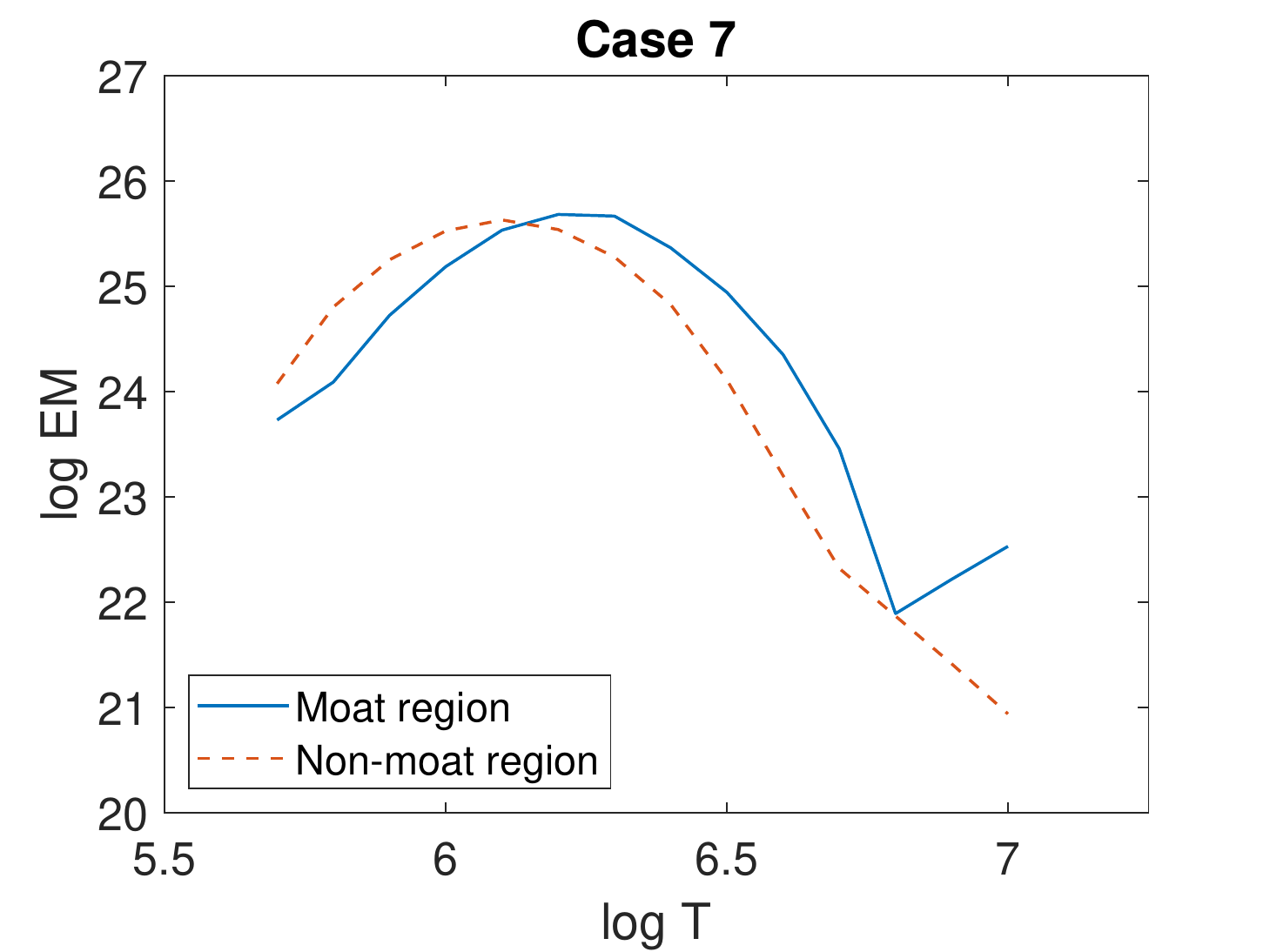} 
\end{tabular}

\caption{Similar to Fig.~\ref{DEM_B_vs_D}, but showing the average log~EM 
calculated for cases 2 to 7, with the method described in Fig.~\ref{DEM_B_vs_D}. 
In all cases, we observe the emission measure in temperature range 
$\sim$log~T=5.8---6.1~MK to be less in the moat-like dark regions around ARs than 
in the surrounding normal-brightness quiet Sun.  EM has the units of cm$^{-5}$ and T has units of Kelvin.}
\label{DEM_B_vs_D_all}
\end{figure}
%\end{comment}

\begin{figure}[!htb]
%\vspace{11pc}
\center
\begin{tabular}{c}  

\includegraphics[scale=0.1,angle=0,height=6cm,keepaspectratio]{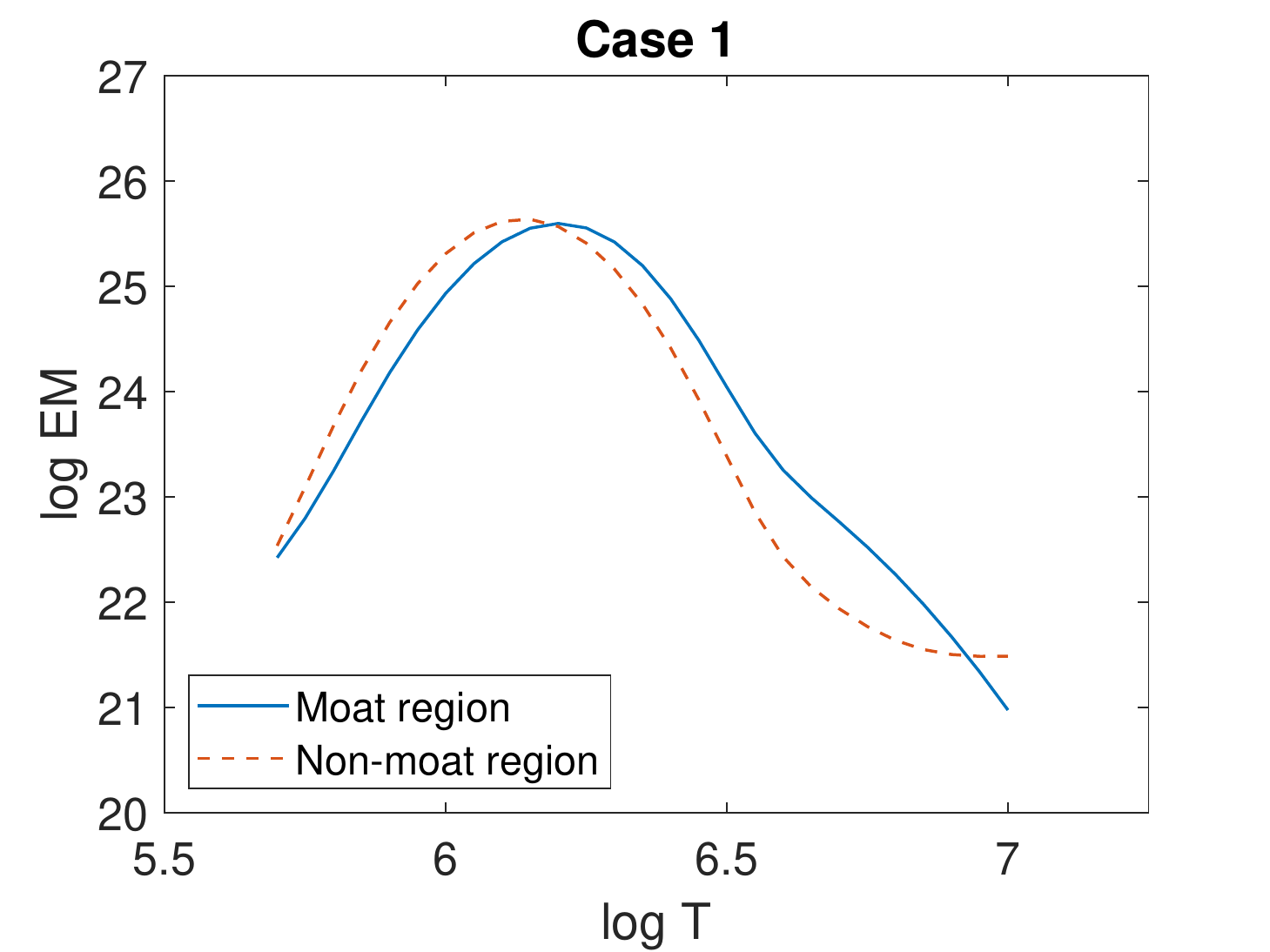} 
%\vspace{11pc}
%\includegraphics[scale=0.1,angle=0,height=6cm,keepaspectratio]{Mark_vs_Markus_10MK_limit.eps}
\end{tabular}

\caption{\textup{Average  log~EM curves calculated
over the dark-region inside the moat boundary but excluding the AR (solid line) and outside the moat boundary but within 60$^o$ of solar disk center (dotted line) for Case 1 using the method described by \citet{Aschwanden13}. EM has the units of cm$^{-5}$ and T has units of Kelvin}}
%(\textit{Right panel}) Comparison of average log~EM curves for regions outside the moat boundary but within 60$^o$ of solar disk center found using the methods described by \citet{Cheung2015} (solid line) and \citet{Aschwanden13} (dotted line).  EM has the units of cm$^{-5}$.}}
\label{Mark_vs_Markus}
\end{figure}

\subsection{Potential Magnetic Field Structure}

To explore the explanation for the presence of these moat regions, we first examine the
magnetic field topology in the regions. We use the Potential Field Source Surface (PFSS)
model \citep{Schatten1969,schrijver.et03}, which uses HMI synoptic magnetograms of the
photospheric magnetic field as an input for calculating  the potential magnetic field in
the corona. The source surface, above which the magnetic field lines  are taken to be
radial, was chosen the be at 2.5~$R_\odot$. 

\begin{figure}[!htb]
%\vspace{11pc}
\center
 
\includegraphics[scale=0.1,angle=0,width=10cm,keepaspectratio]{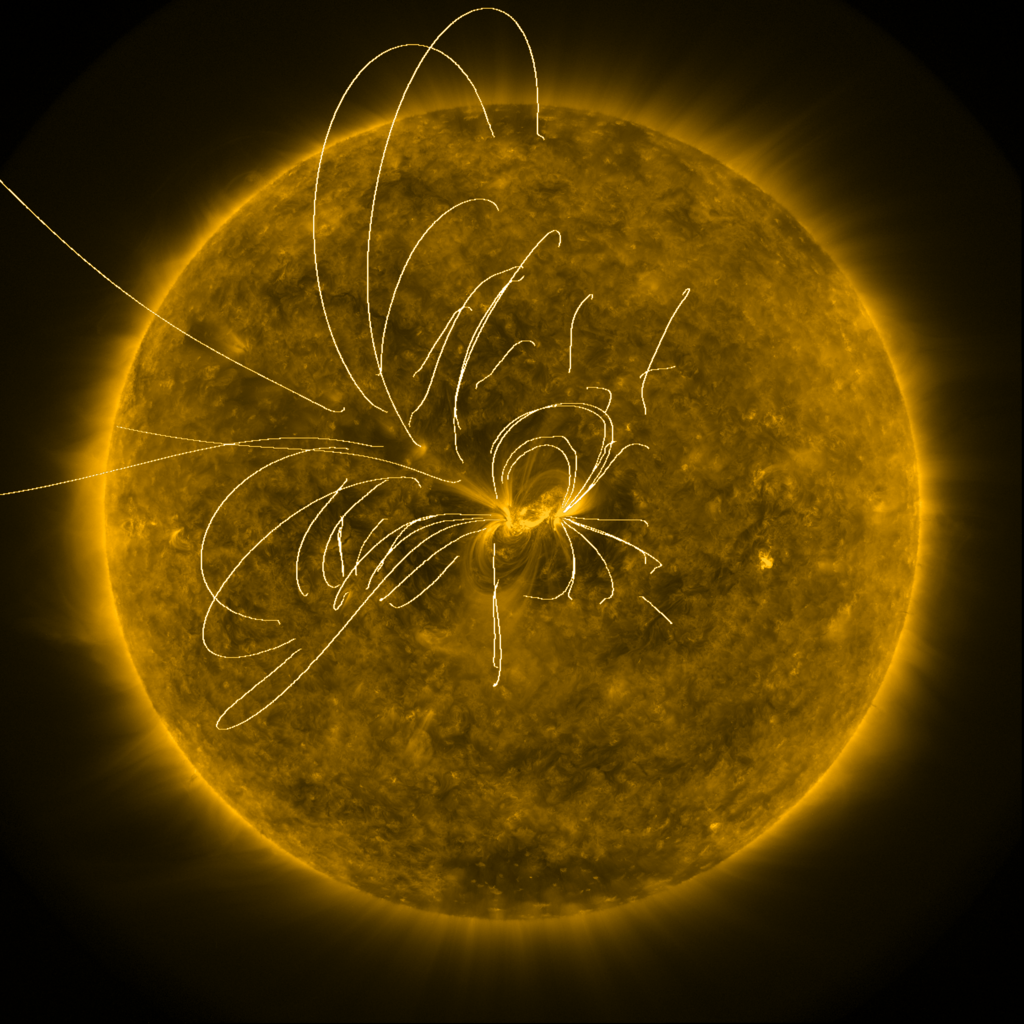}

\caption{\sdo/AIA 171~{\AA} image of Fig.~\ref{Without_PFSS} (2018 Nov 11, 06:04~UT), 
for Case~1 of Table~1, over-plotted with potential field lines generated 
from the PFSS lines.  We display 
only field lines originating in or near the dark moat region around AR 12699.}
\label{PFSS}
\end{figure}

Figure~\ref{PFSS} shows the magnetic field lines obtained using the PFSS model for
case~1,  over-plotted over the 171~{\AA} image of Figure~\ref{Without_PFSS}. \textup{We have used the $pfss\_viewer$ package in $SolarSoft$ to obtain the PFSS solution here.} We only show
the field  lines that start or end in or near the dark moat. We find various
resulting  magnetic-field-line connections originating in the moat: from the moat to a
sunspot in the AR, from the moat to bright quiet-Sun locations, and also field
lines rooted in the moat that open out into the heliosphere.  Thus we cannot
identify any special outstanding characteristic of the potential field rooted 
in the moat surrounding the AR\@. \textup{We have tried plotting much larger numbers of lines in the vicinity of the AR, but no special characteristic could be identified for lines rooted in the moat.}

Because we do not see any systematic pattern in the resulting magnetic field topology, we
conclude  that the PFSS solution alone is insufficient for explaining the origin of these
AR-surrounding dark moats.  Moreover, the presence of closed loops with one end rooted in
the dark moat reaffirms that the moat is not a coronal-hole region.

%This might be due to the PFSS assumption that the magnetic field is potential; that assumption is 
%likely insufficient in this case. 

%\begin{comment}
%\end{comment}

\subsection{Possible Cause: Flat Loops}\label{monopole}

In this section, we consider a different possible explanation for the dark-moat regions, one that relies
on the coronal loops in the outskirts of the AR being forced to be flat (nearly horizontal to
the  solar surface) by the pressure of the overlying strong magnetic field of the AR\@.  

This is the background to our concept. \citet{Antiochos86} explain the structure of the
static corona by pointing out that there are two possible models for coronal loops; that
is, the coronal-loop energy-balance equations admit two  different possible solutions. 
These two solutions are: (1) a hot-loop model and (2) a cool-loop model; respectively yielding loops
of temperature $>10^5$~K, and between about  $2\times10^4$ and $1\times 10^5$~K. They argued that the  hot solution is only available for loops that reach
high-enough heights compared to the local coronal density scale height.  Specifically,
they say that the hot solution only works on loops with  heights of $\gtsim$5000~km;  the
hot solution becomes thermally unstable for loops having top heights lower than $5000$~km.  
Consequently,  if the loops have top heights less than $\sim$5000~km they are restricted to
temperatures only  in the range of about $2\times10^4$---$1\times 10^5$~K\@. Thus, in the
general solar atmosphere, all of the one-million- and multi-million-K hot coronal loops
are those that have top heights in excess of $\sim$5000~km. This theoretical result is key to our suggested explanation for the dark moats surrounding
ARs.

Because the moats have a propensity to be dark in only the cooler-temperature AIA
channels, we suspect that coronal loops that would normally emit largely in the
temperature range  $\sim$0.5---1.0~MK get flattened to  heights less than $\sim$5000~km,
meaning that the hot solution stops working for them and only the cool solution of
\citet{Antiochos86} is available. Such low-lying loops therefore would not emit strongly 
at wavelengths from plasma in that temperature range.  Thus, if the loops in the outskirts of ARs that would normally emit in this
$\sim$0.5---1.0~MK range, in particular the loops that would normally emit in the AIA~171~\AA\ band, are flattened to
low heights, then this could explain why such regions are dark in images taken with
filters sensitive to such wavelengths. Such flattened loops would tend to have temperatures below
$10^5$~K, and therefore they would tend to be dark in emission from plasma that emits in the AIA~171~\AA\ band.

We suspect that this is the explanation for the moat regions.  Because the loops that
tend to emit in 171~\AA\ in the normal quiet Sun (e.g., non-moat quiet Sun) are those
that tend not to be too long \citep[this follows from the loop-scaling law,
e.g.,][]{rosner.et78}, such loops would, on average, have footpoints that are not far
separated compared to the size of the AR\@. We expect that the overlying magnetic field of the adjacent AR imparts a downward
magnetic pressure high enough to confine such relatively short loops in the outskirts of
the AR, along with the lowest of the AR's loops that splay out from the AR, to low enough
heights to bring their temperatures below $\sim$10$^5$~K\@. 

To examine the plausibility of this claim, we estimate the region around the ARs where the field \textup{pressure} of the overlying splay of magnetic field from the AR is sufficient to suppress the heights of coronal loops in those surrounding regions.  In order to suppress those surrounding loops, \textup{the magnetic pressure of the AR's outskirt field over the moat regions should be higher than beyond the non-moat regions and the moat boundaries should mark where the magnetic pressure has weakened to comparable to the magnetic pressure at the base of the corona in the quiet sun regions outside the moat.}   

\begin{figure}[!htb]
%\vspace{11pc}
\center
\begin{tabular}{c c}  
\hspace{-0.6pc}
\begin{overpic}[scale=0.1,angle=0,width=7cm,keepaspectratio]{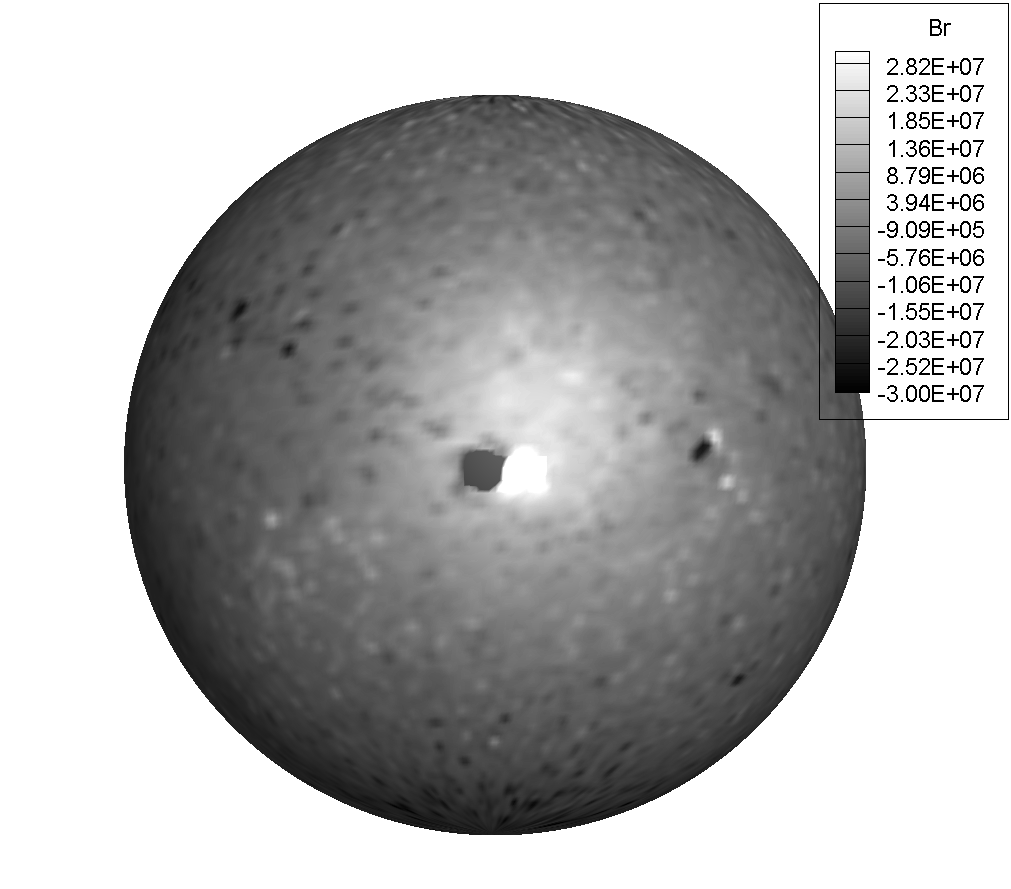}
\end{overpic}
\begin{overpic}[scale=0.1,angle=0,width=7cm,keepaspectratio]{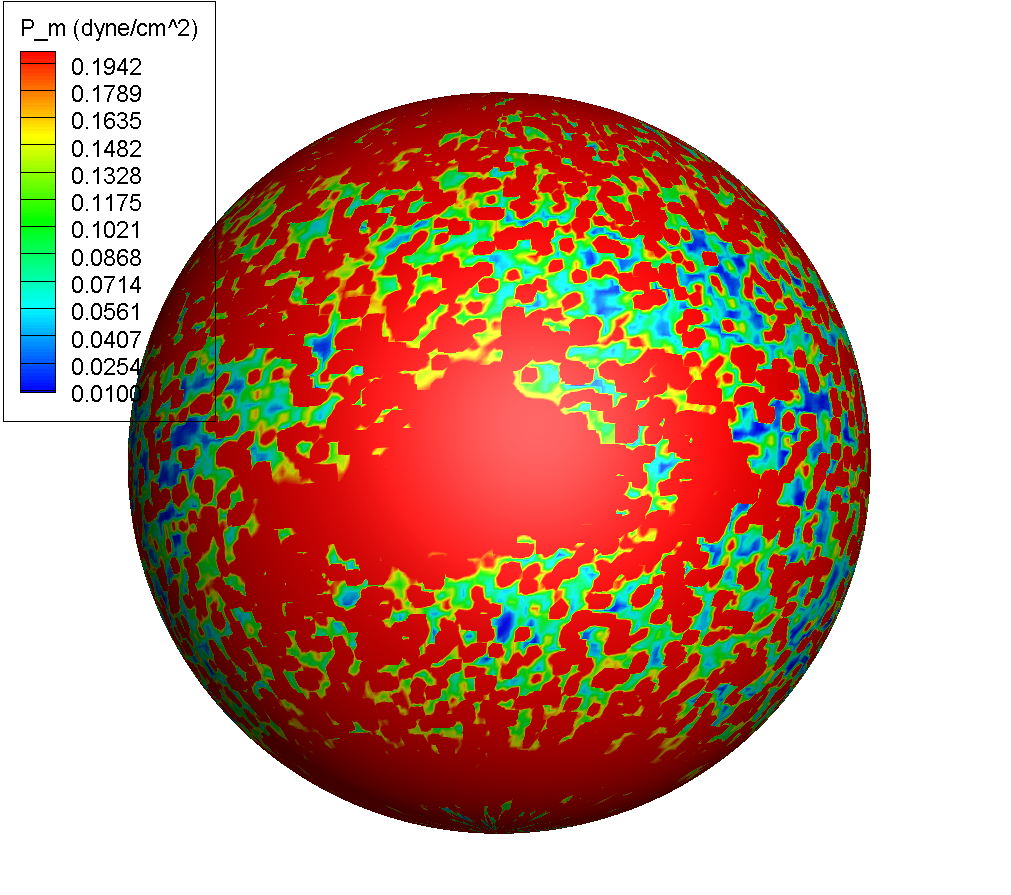}
\end{overpic}\\
\begin{overpic}[scale=0.1,angle=0,width=7cm,keepaspectratio]{1-AIA_171_dark_Region_shrink.jpg}
\end{overpic}
\begin{overpic}[scale=0.1,angle=0,width=7cm,keepaspectratio]{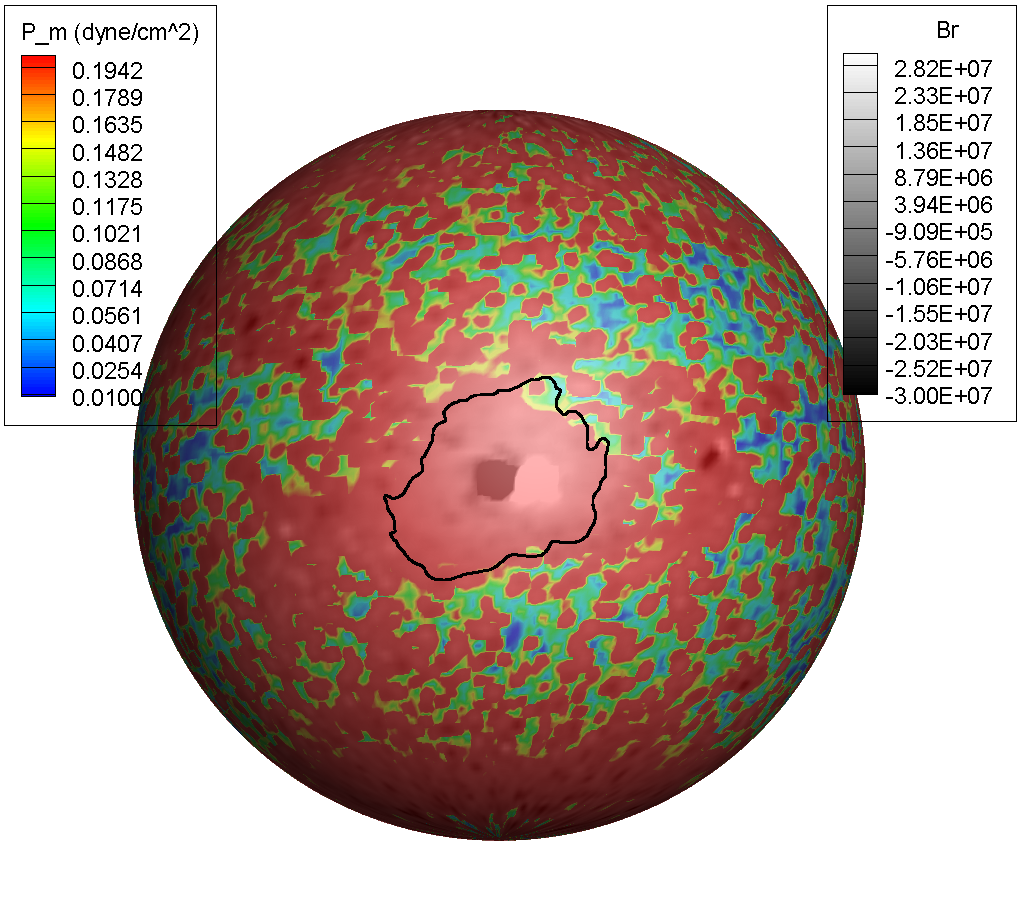}
\end{overpic}

\end{tabular}

\caption{\textup{\textit{(Top-Left)}: The radial synoptic magnetogram  for case 1, used to calculate the PFSS solution. The units of radial magnetic field ($Br$) in this graph are $\mu$Gauss. \textit{(Top-Right)}: The magnetic pressure ($P\_m$) found using the PFSS solution at 1.03 $R_\odot$. Since the upper limit of the $p\_m$ colorbar is chosen as 0.2 $dyne/cm^2$, the red color represents magnetic pressure near or exceeding 0.2 $dyne/cm^2$. The maximum magnetic pressure was over the AR and was 7162 $dyne/cm^2$. \textit{(Bottom-Left)}: The AIA 171 \AA\ image for case 1, with the visually drawn moat black outer boundary. \textit{(Bottom-Right)}: The spheres from top two panels are plotted together. The magnetogram sphere is visible below the pressure sphere because we made the pressure sphere semi-translucent. The moat boundary shown in bottom-left panel is over-plotted on the pressure sphere. We notice that the quiet sun has regions of magnetic pressure around or below 0.2 $dyne/cm^2$, but the region in the vicinity of the AR has magnetic pressure exceeding 0.2 $dyne/cm^2$. The moat boundaries roughly lie where we start observing regions with magnetic pressure less than 0.2 $dyne/cm^2$, while moving away from the AR.}}
\label{contour1}
\end{figure}

\textup{The first step then is to determine the magnetic pressure at the coronal-base heights everywhere on the Sun. We do this by using the HMI synoptic magnetograms to create PFSS solutions using the Finite Difference Iterative Potential-field Solver (FDIPS) for all seven cases in Table~\ref{table1st} \citep{Toth16}. We then calculate the magnetic pressure using this PFSS solution on a surface with radius $1.03R_\odot$ (20,000 km above the photosphere). The synoptic magnetogram for Case~1 and the corresponding PFSS pressure at $1.03R_\odot$ is shown in the top panels of Fig.~\ref{contour1}. In the bottom-right panel of Fig.~\ref{contour1}, the pressure surface has been made semi-translucent so that the synoptic magnetogram used to obtain the PFSS solution can be seen below it on the spherical surface of the Sun's photosphere. Next, we overplot the moat boundary found visually from the corresponding AIA 171 \AA\ image (bottom-left panel of Fig.~\ref{contour1}) on the magnetic pressure surface. We have saturated the colorbar so that the magnetic pressure distribution on non-moat regions is clearly visible. The quiet sun has magnetic pressure randomly distributed between 0.01 and 0.2 $dyne/cm^2$, but the region in the vicinity of the AR has magnetic pressure exceeding 0.2 $dyne/cm^2$. The moat boundaries roughly lie where we start observing regions with magnetic pressure less than 0.2 $dyne/cm^2$, as we move away from the AR.}

Figure~\ref{contour1} shows our results for this exercise for the AR of case~1 of
Table~1. \textup{ We can see that the moat boundary encapsulates a high magnetic pressure region, relative to the non-moat regions. Although the boundaries do not match exactly, there is obvious correlation between the dark moat region in AIA 171 \AA\ image and the high pressure region in the PFSS solution. This point can be further validated by looking at the rest of the cases.} 

\begin{figure}[!htb]
%\vspace{11pc}
\center
\begin{tabular}{c c}  
\hspace{-0.6pc}
\begin{overpic}[scale=0.1,angle=0,width=6cm,keepaspectratio]{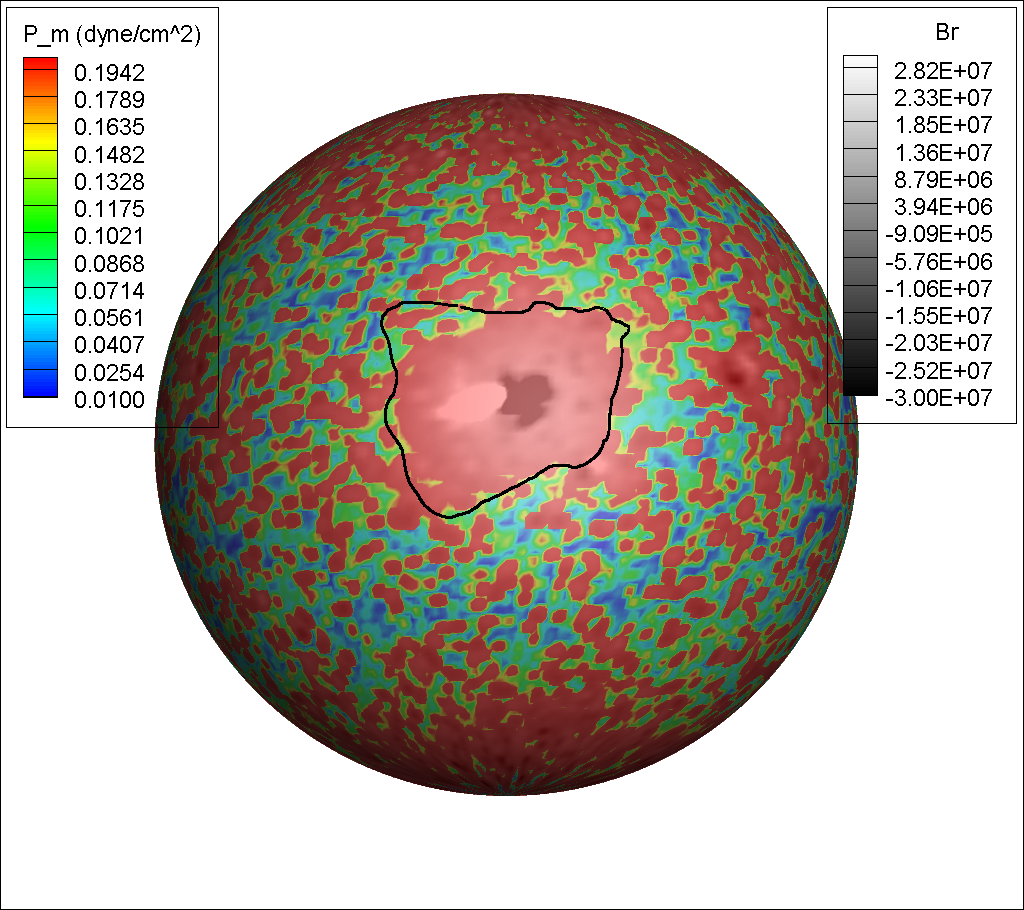}
\put(5,5){\color{black}{ \fontsize{8}{9}\selectfont 2}}
\end{overpic}
\begin{overpic}[scale=0.1,angle=0,width=6cm,keepaspectratio]{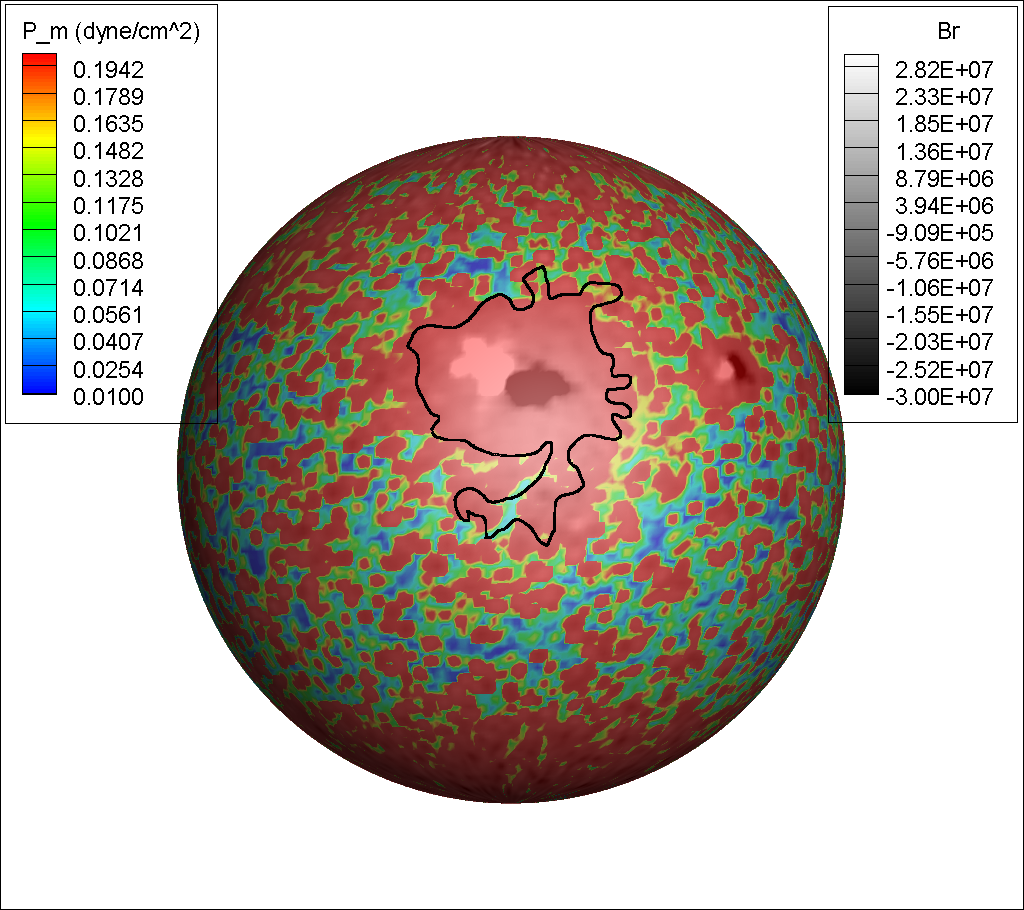}
\put(5,5){\color{black}{ \fontsize{8}{9}\selectfont 3}}
\end{overpic}\\

\hspace{-0.6pc}
\begin{overpic}[scale=0.1,angle=0,width=6cm,keepaspectratio]{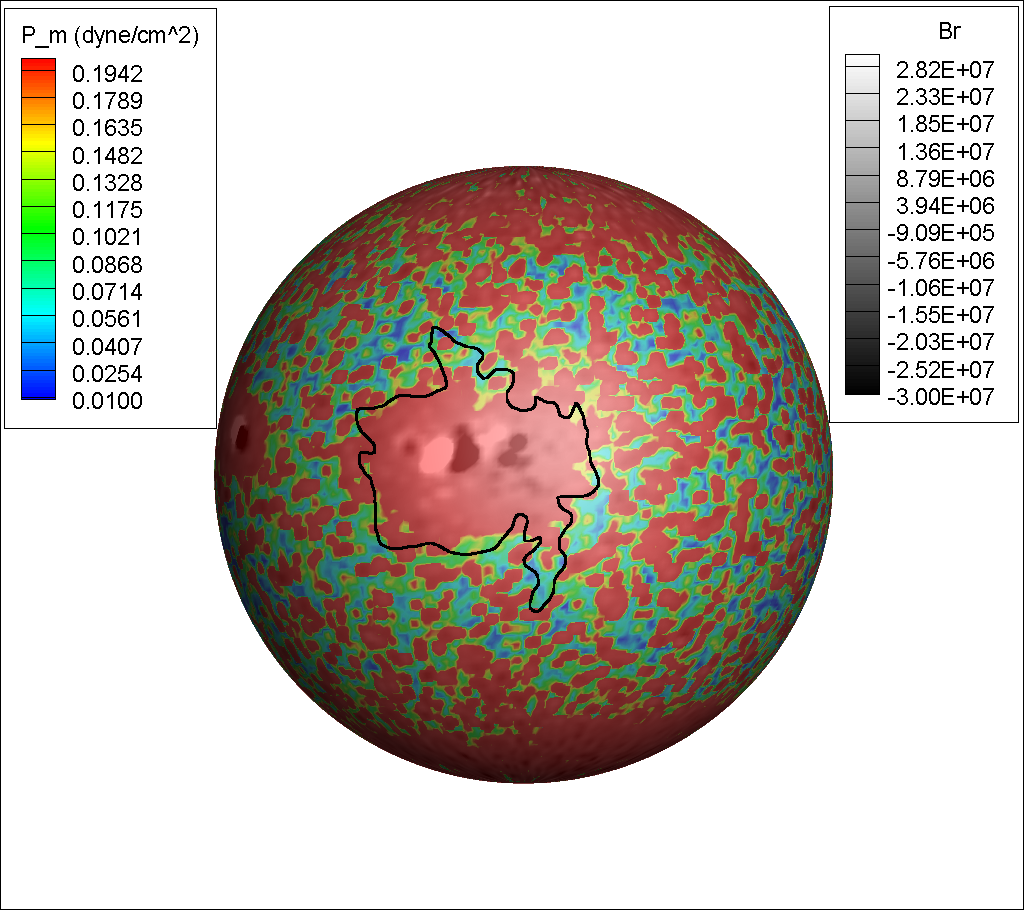}
\put(5,5){\color{black}{ \fontsize{8}{9}\selectfont 4}}
\end{overpic}
\begin{overpic}[scale=0.1,angle=0,width=6cm,keepaspectratio]{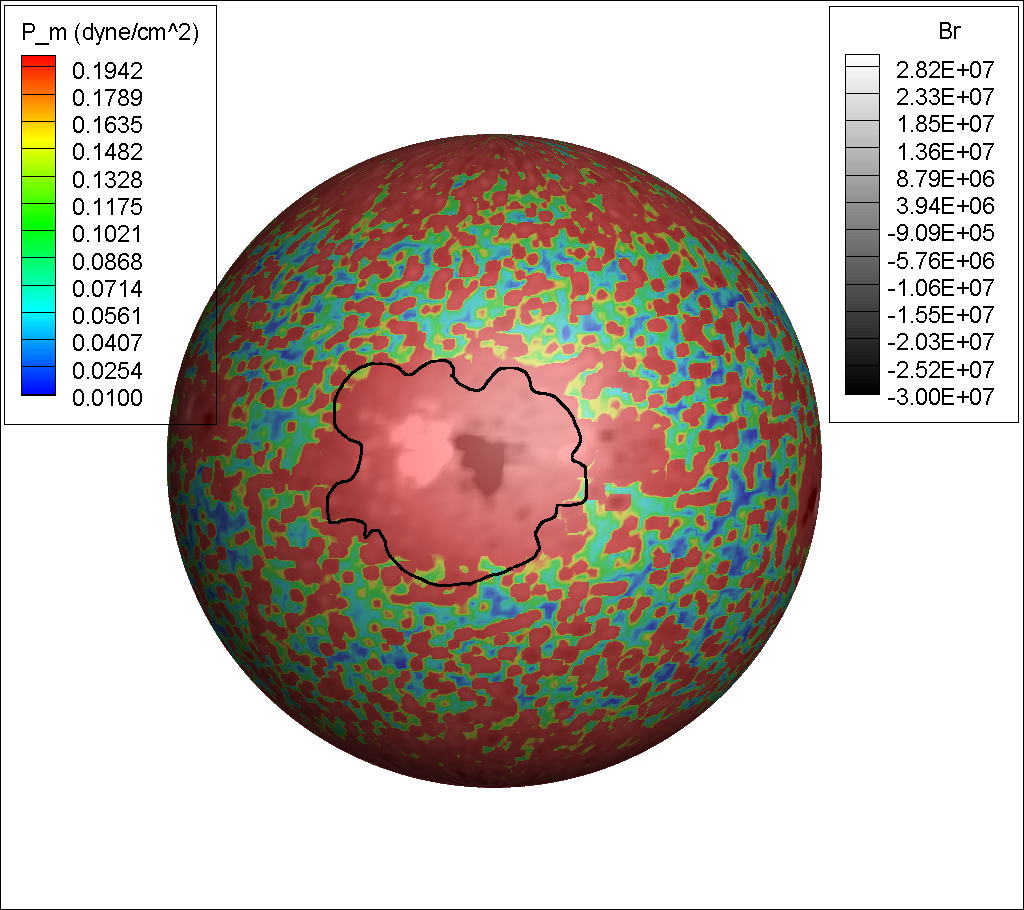}
\put(5,5){\color{black}{ \fontsize{8}{9}\selectfont 5}}
\end{overpic}\\

\begin{overpic}[scale=0.1,angle=0,width=6cm,keepaspectratio]{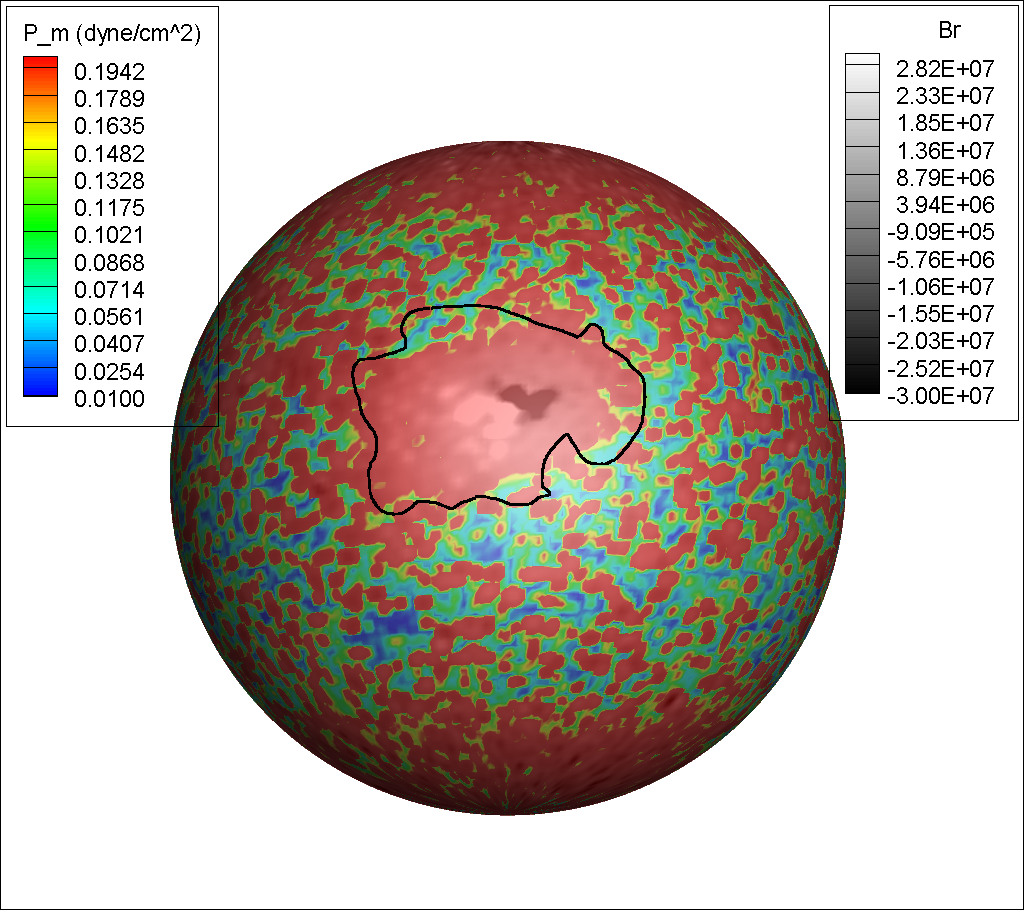}
\put(5,5){\color{black}{ \fontsize{8}{9}\selectfont 6}}
\end{overpic}
\begin{overpic}[scale=0.1,angle=0,width=6cm,keepaspectratio]{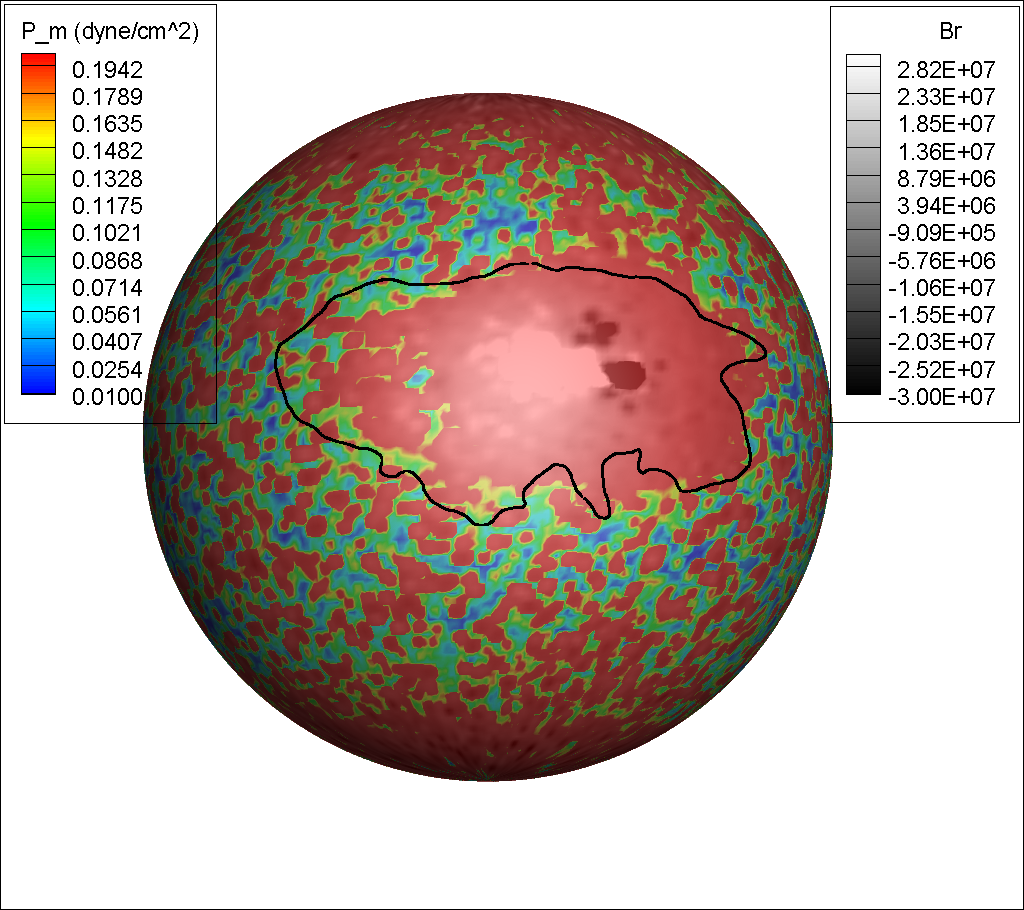}
\put(5,5){\color{black}{ \fontsize{8}{9}\selectfont 7}}
\end{overpic}

\end{tabular}

\caption{\textup{The magnetic pressure ($P\_m$) found using PFSS solution at 1.03 $R_\odot$ is shown using a semi translucent sphere for cases 2--7 in Table~\ref{table1st}. There is another sphere below this showing the synoptic magnetogram used to calculate the PFSS solution. The units of radial magnetic field ($Br$) in this graph are $\mu$Gauss. The visually found moat boundaries shown in Fig.~\ref{all171} are over plotted on this sphere. The quiet sun magnetic pressure seems randomly distributed between 0.01 and 0.2 $dyne/cm^2$, but the region in the vicinity of the AR has magnetic pressure exceeding 0.2 $dyne/cm^2$. The moat boundaries roughly lie where we start observing regions with magnetic pressure less than 0.2 $dyne/cm^2$, while moving away from the AR.}}
\label{contour_all}
\end{figure}

Figure~\ref{contour_all} shows contours made in the same fashion
for  cases~2---7 of Table~1.  \textup{In each of the cases the results are similar, in that the moat boundaries found using AIA 171 \AA\ data again roughly bound the region of high magnetic pressure around the AR. The minor mismatches might be due to our potential field approximation of the coronal magnetic field or errors in visual estimation of the moat boundaries from AIA 171 \AA\ data.}

Overall however, we regard the match as good enough to render
plausible the  idea that a moat region around ARs is often dark in 171~\AA\ and similar
wavelengths because the strong AR field pushes down the coronal loops in that region that
otherwise would emit strongly in 171~\AA\@.  Those low-lying coronal loops do not emit in
171~\AA\ because hot emission  in loops of any length confined to low-enough heights is not allowed due to the thermal
instability pointed out by \citet{Antiochos86}.

%\begin{comment}
%\end{comment}
\section{Summary and Discussion}\label{conclusions}

We have re-examined the dark moats frequently seen to surround ARs in AIA EUV coronal images, especially
in emission from plasma at  temperature near 1~MK\@.   In our seven selected cases, we found that
these dark moats vary in size and shape, but are present in
all of our cases. Using DEM analysis, we found that in these moat regions, there is a dearth of
emission at temperatures centered around \textup{0.6---1.1~MK}\@.  Since the 171~{\AA} channel of AIA is most
sensitive in this temperature range, the dark moats are most pronounced in the AIA 171~{\AA} images. We performed PFSS magnetic field analysis in the moat regions to find whether
there was any systematic pattern for the dark moats, but we could not find any
special magnetic topology of field lines rooted in the moat areas to explain the reduced emission.

\textup{By looking at the magnetic pressure distribution around the ARs using PFSS solutions}, we find plausible the idea that these
dark moats are a collateral consequence of the splay of strong loop magnetic field that
extends out from the strong and concentrated magnetic flux of the AR\@.  We argue that
the overlying strong field suppresses the maximum height attained by medium-length coronal loops
in those surrounding regions (where here by ``medium-length" we mean those that normally 
would emit in this near-1-MK temperature window, in accordance to loop length-temperature
scaling law of \citet{rosner.et78}), and also keep the AR's own
lowest loops that low over the moat area.  According to \citet{Antiochos86}, those
height-suppressed medium-length loops, and the overlying AR-rooted lowest loops, could
only achieve temperatures of  $\sim$10$^5$~K, due to thermal instability of the plasma in them that
ensues at higher temperatures.

The higher AR-rooted loops that stretch out over and above those low-lying moat loops
apparently achieve  high-enough heights that they are not subject to the high-temperature 
thermal instability. Consequently, their peak temperatures are in excess of the near-1-MK normal temperature of unflattened medium-sized loops.  Their temperature and consequent plasma density and EUV emission are comparable to or greater than that of the quiet-region corona, plausibly due to their obeying the loop scaling law and – from being rooted in the strong magnetic flux of the AR, which gives these loops heating comparable to or greater than that of the quiet-region corona. Apparently, their heating is strong enough that they are heated to temperatures above 1 MK.  This
would explain why the moats that are dark in AIA 171~\AA\ images are reduced in prominence or
absent (i.e., the brightness of the AR-outskirt regions is comparable to the non-moat
quiet regions) in AIA filters sensitive to higher-temperature  emissions (Fig.~\ref{case1}).

\citet{wang11} analyzed these dark moats in 171~{\AA}, and concluded that they consist of
171~\AA-EUV absorbing ``fibrils'' that might be the coronal counterparts of the fibrils
as seen in the H$\alpha$ chromosphere data. They argue that the \halpha\ fibril fields at
the bottom of the corona reconnect with encountered overlying field from the bipolar AR,
resulting in a fibril-to-filament conversion, and those filaments  constitute the dark moats.  Their EUV-absorbing-fibril concept might complement our idea for the dark moats, as the low-lying cool coronal loops, expected to be of temperature $\ltsim$10$^5$~K, would likely absorb any 171~\AA\ emission from below them, if any.

\begin{figure}[!htb]
%\vspace{11pc}
\center
 
\includegraphics[scale=0.1,angle=0,width=10cm,keepaspectratio]{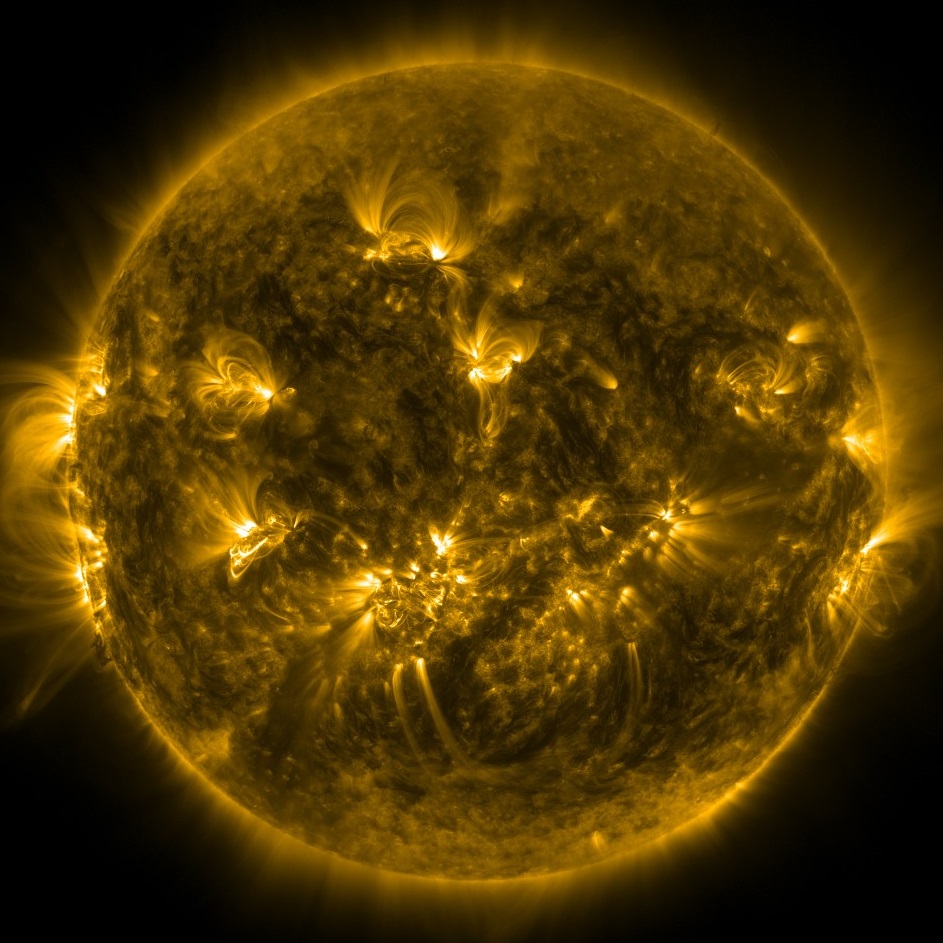}

\caption{An \sdo~AIA~171~{\AA} image from 2014 Dec~17, 03:12~UT\@. This was 
during solar maximum, while all of the examples of Table~1
are from solar minimum.  Moat-like dark regions are visible around ARs here also,
and, at latitudes below $\pm60^o$, they cover a significant portion of the solar disk owing to the presence of numerous 
ARs.  Thus the moat regions occur independent of whether there is only one AR on the 
disk during minimum, as in the case of the Table~1 events, or when multiple ARs
are concurrently on the disk during solar maximum.}
\label{maxima}
\end{figure}

For this study we selected seven cases when there was just one AR on the solar disk to
focus our efforts, and all of our examples ended up being from during solar minimum. 
\textup{Our inspections of AIA data at different times over the solar cycle since SDO's launch in 2010 suggest that} these dark moat regions are present near ARs even when there are multiple ARs on
the Sun, and throughout the  solar cycle.  Figure~\ref{maxima} shows an AIA 171~\AA\
example from 2014~Dec~17, during solar maximum.  We can see that low-emission dark regions (similar in darkness to the moats of our lone ARs) cover the entire longitude
range at AR latitudes and somewhat higher latitudes.  Therefore we find the dark moats around ARs in
the near-1-MK temperature range to be a general phenomenon. Here we have argued that the
dark moat regions are a consequence of the strong peripheral AR fields pushing down on
surrounding medium-sized coronal loops, and pushing down on the bottom-most of the AR's
own peripheral loops, so that those loops are low-lying enough to be restricted to
lower-than-coronal temperatures, following \citet{Antiochos86}.  This effect, combined
with the presence of overlying hotter loops rooted in the AR, results in a depletion of cool-corona plasma in the outskirts (dark moats) of ARs.  This plausibly
explains why the dark moat regions are seen around ARs in cool-corona EUV images.

\textup{An unresolved aspect of our suggested explanation for the moat regions is that \citet{Antiochos86} assume that coronal loops are static, while many observations indicate that AR coronal loops are dynamically changing \citep[e.g.,][]{Warren03, Winebarger03}. Those observations however have tended to focus on loops within active regions,
rather than in the periphery of the active regions where the moats form.  Thus to address the
question of how dynamic motions on loops might alter our suggested mechanism for the moats,
both the nature of the dynamics of loops in the moat regions should be examined 
observationally, and the consequences of the dynamic behavior of loops on loop
thermal structure should be studied with numerical simulations.  To our knowledge, such studies
have not yet been carried out to address the specific question of EUV intensity in the moat regions.}

\textup{Another unresolved aspect of our suggested explanation is the appearance of the moats in 304 \AA\ images.  Because the 304 line is formed at about $5 \times 10^4$ K \citep{ODwyer10}, 
it might be expected that the cool coronal loops, being of temperature less than $5 \times 10^5$ K, might 
cool into the emission passband of the 304 \AA\ channel, and therefore appear bright in 
304 \AA\ images.  Yet the moats are generally visible as darker areas in 304 \AA\ images \textup{(we only 
make this determination visually, since the DEM procedure of \S\ref{DEM_section} is only 
applicable down to $T\sim 5\times 10^5$\,K)}\@. 
A possible explanation for this could be that material making up the dark fibrils 
discussed by \citet{wang11} could block and scatter some of the lower-altitude 304 \AA\ 
emission, leading to the reduced 304 intensity in the moats.  Future spectroscopic studies
of moat regions should be beneficial in understanding this and other aspects of the moat 
regions.}

%\begin{comment}
%\end{comment}

We thank M. Chueng for clarifications regarding the \citet{Cheung2015} code.  We
also thank the referee for suggestions that greatly improved the paper.  A.C.S. and 
R.L.M. received funding from the  Heliophysics Division of NASA's Science Mission
Directorate  through the Heliophysics Guest Investigators (HGI) Program.

\appendix
\section{Emission measure calculation}\label{appendix_A}

\textup{\citet{Cheung2015} derive the EM in 21 sequential temperature intervals 
by setting up and solving a severely underdetermined system of linear 
equations.  They set up the system by dividing the $\log T$ range into 21 
intervals, covering 5.5 to 7.5 in increments of 0.1 in $\log T$\@.  They place 
four mathematical functional shapes in each of the 21 intervals, each 
representing some of the DEM in that temperature interval.  One functional 
form is a delta function, with a constant contribution to the DEM at all 
temperatures in that interval, and that is done for each of the 21 intervals.  
The other three functional forms are Gaussian (of three differing widths), 
but where the Gaussians are truncated so that they do not have infinitely 
long tails.  Thus, in total, they represent the total EM in the entire temperature 
range by a sum over these $21\times 4 = 84$ different DEM functional forms, where each 
of the 84 DEMs is weighted by an unknown coefficient.  In the case of the AIA 
data, there are six intensities, from the six AIA coronal EUV channels, that can 
be measured at each pixel in the AIA field of view.  This means that, for each pixel, 
there are six measurements with 84 unknowns, resulting in the severely 
underconstrained mathematical problem.  \citet{Cheung2015} proceed to 
solve this mathematical problem by utilizing techniques for underconstrained 
systems from the field of ``compressed sensing.''   They use an approach 
called ``sparsity'' \citep[e.g.,][]{candes.et06,candes.et07}, which includes imposing 
constraints on the underconstrained system. The \citet{Cheung2015} constraints 
include requiring all of the 84 DEM coefficients to have non-negative values, 
that the sum of those coefficients be minimized, and that for each of the six AIA channels the predicted count 
rate (DN/s) from the model DEM be within some tolerance of the observed count 
rate.  The solution they arrive at gives them the 
EM in each of the 21 bins at each pixel.  \citet{Cheung2015} 
then proceed to show that this approach produces satisfactory results using 
both synthetic and actual AIA data.  See \citet{Cheung2015} for more details.}

\clearpage
%\bibliography{sample63}
%\bibliography{mcc} %Talwinder commented
%\bibliographystyle{aasjournal} %Talwinder commented

\end{document}